\documentclass[10pt, journal ]{IEEEtran}

\usepackage{amsthm}
\usepackage{amsfonts}
\usepackage{amssymb}
\usepackage{mathrsfs}
\usepackage{mathtools}
\usepackage{float}
\usepackage[pdftex]{graphicx}
\usepackage{amsmath}
\usepackage{color}
\usepackage{multirow,multicol}
\usepackage{graphicx}
\usepackage{times}
\usepackage{textcomp}
\usepackage{verbatim}
\usepackage[table]{xcolor}
\usepackage{balance}
\usepackage{lipsum}
\usepackage[inline]{enumitem}
\usepackage{cuted}
\usepackage[caption=false,font=footnotesize]{subfig}
\usepackage{cite}
\usepackage{algpseudocode,algorithm}
\usepackage{hyperref}
\usepackage{tcolorbox}
\usepackage{setspace,epstopdf}
\usepackage[table]{xcolor}
\definecolor{color1}{RGB}{199,209,232}
\definecolor{color2}{RGB}{230,231,233}

\DeclareMathOperator*{\maximize}{maximize} 
\DeclareMathOperator*{\minimize}{minimize} 
\DeclareMathOperator*{\subjectto}{subject\hspace{3pt} to:\hspace{3pt}} 
\newtheorem{theorem}{Theorem}

\begin{document}
	
	\title{Antenna Selection With Beam Squint Compensation for Integrated Sensing and Communications}
	

	%

	\author{\IEEEauthorblockN{Ahmet M. Elbir, \textit{Senior Member, IEEE},  Asmaa Abdallah, \textit{Member, IEEE}, \\ Abdulkadir Celik, \textit{Senior Member, IEEE}, and Ahmed M. Eltawil, \textit{Senior Member, IEEE}  }
		
		\thanks{A. M. Elbir, A. Abdallah, A. {C}elik and A. M. Eltawil are with King Abdullah University of Science and Technology, Thuwal 23955, Saudi Arabia (e-mail: ahmetmelbir@ieee.org, asmaa.abdallah@kaust.edu.sa, abdulkadir.celik@kaust.edu.sa, ahmed.eltawil@kaust.edu.sa). } 
		
	}
	
	\maketitle


	\begin{abstract}
		Next-generation wireless networks strive for higher communication rates, ultra-low latency, seamless connectivity, and high-resolution sensing capabilities. To meet these demands, terahertz (THz)-band signal processing is envisioned as a key technology  offering wide bandwidth and sub-millimeter wavelength. Furthermore, THz integrated sensing and communications (ISAC) paradigm has emerged jointly access spectrum and reduced hardware costs through a unified platform. To address the challenges in THz propagation, THz-ISAC systems employ extremely large antenna arrays to improve the beamforming gain for communications with high data rates and sensing with high resolution.  However, the cost and power consumption of implementing fully digital beamformers are prohibitive. While hybrid analog/digital beamforming can be a potential solution, the use of subcarrier-independent analog beamformers leads to the beam-squint phenomenon where  different subcarriers observe distinct directions because of adopting the same analog beamformer across all subcarriers.  In this paper, we develop a sparse array architecture for THz-ISAC with hybrid beamforming to provide a cost-effective solution. We analyze the antenna selection problem under beam-squint influence and introduce a manifold optimization approach for hybrid beamforming design. To reduce computational and memory costs, we propose novel algorithms leveraging grouped subarrays, quantized performance metrics, and sequential optimization. These approaches yield a significant reduction in the number of possible subarray configurations, which enables us to devise a neural network with classification model to accurately perform antenna selection. Numerical simulations show that the proposed approach exhibits up to 95\% lower complexity for large antenna arrays while maintaining satisfactory communications with approximately 6\% loss in the achievable rate.

	\end{abstract}

	%
	\section{{Introduction}}
	
	\label{sec:Introduciton}
	\IEEEPARstart{T}{he} {escalating demand for wireless communications and radar systems has engendered a scarcity of available frequency bands, resulting in pervasive overcrowding and spectrum congestion~\cite{mimoScalingUp,mishra2019toward}. To combat this predicament,  specialized techniques such as carrier aggregation and spectrum stitching have been harnessed in communications systems to efficiently utilize the spectrum \cite{heath2016overview}. However, the application of these techniques to radar systems poses formidable challenges in achieving meticulous phase synchronization~\cite{mishra2019toward}. Consequently, it is crucial to cultivate approaches that enable the simultaneous and opportunistic operation within the same frequency bands, thus benefiting from both radar sensing and communications functionalities on a shared hardware platform. Therefore, there has been a significant interest focused on the development of strategies for integrated sensing and communications (ISAC) setups, aiming to jointly access the scarce spectrum in a mutually advantageous manner \cite{elbir_thz_jrc_Magazine_Elbir2022Aug,jrc_TCOM_Liu2020Feb,elbir2021JointRadarComm}.} 
	

	\begin{figure*}[t]
		\centering
		{\includegraphics[draft=false,width=.95\textwidth]{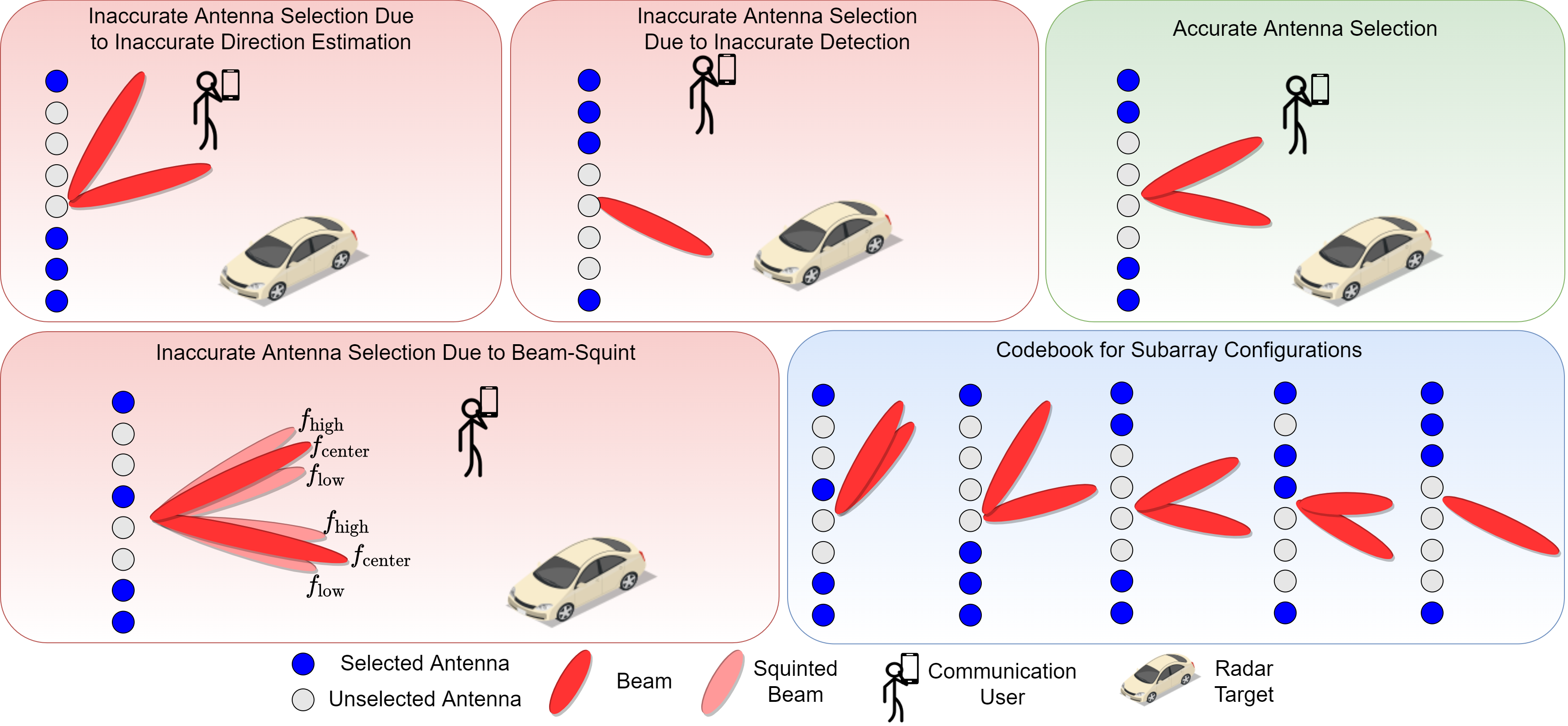}} 
		\caption{Antenna selection for various scenarios. A codebook of subarray configurations is given to accurately select the antennas. Inaccurate antenna selection emerges  due to failed direction estimation, misdetection and beam-squint which causes the generated beams point to different directions across the subcarriers.
		}
		\label{fig_AntennaSelectionDiag}
	\end{figure*}

	{The earlier ISAC designs utilize distinct hardware platforms to carry out sensing and communications (S\&C) functions within the same frequency bands. These designs employed various techniques to mitigate interference between the two domains.  Broadly, the ISAC systems are categorized into two primary groups: radar-communications coexistence (RCC) and dual-functional radar-communications (DFRC) \cite{mishra2019toward, fanLiu_JRC_JSTSP_Zhang2021Sep}. While RCC focuses on managing interference and sharing resources between S\&C tasks, enabling them to operate without significant mutual disruption, DFRC aims to consolidate both tasks onto a common platform, resulting in the convergence of ISAC design \cite{elbir_thz_jrc_Magazine_Elbir2022Aug,fanLiu_JRC_JSTSP_Zhang2021Sep}. The necessity for a unified hardware platform becomes increasingly imperative as the integration of communications and sensing capabilities continues to advance in various applications, such as vehicle-to-everything (V2X) communication, indoor localization, radio frequency (RF) tagging, extended/virtual reality, unmanned aerial vehicles (UAVs), and intelligent reflecting surfaces (IRSs) \cite{elbir2021JointRadarComm, thz_isac5_Petrov2019May, heath_Sensing4Comm_Choi2016Dec, localization_ISAC_Wymeersch2022Jun, thz_2030_Chen2021Nov, iot_ISAC_Cui2021Nov, seven_THZ_ISAC_walid2021SensingComm, thz_isac3_Chang2022Mar, elbir_IRS_ISAC_Elbir2022Apr}.  The terahertz (THz) band ($0.1-10$ THz) has emerged as a promising technology to meet the sixth-generation (6G) wireless networks' ambitious performance goals on enhanced mobile broadband (eMBB), massive machine-type communication (mMTC), and  ultra-reliable low-latency communication (URLLC) \cite{thz_2030_Chen2021Nov, elbir_thz_jrc_Magazine_Elbir2022Aug}. As the spectrum allocation beyond 100 GHz is underway, there is a surge of research activity in ISAC to develop system architectures that can simultaneously achieve high-resolution sensing and high data rate communication at both upper millimeter-wave (mmWave) frequencies and low THz frequencies \cite{mishra2019toward, thz_isac5_Petrov2019May, elbir2021JointRadarComm, elbir_thz_jrc_Magazine_Elbir2022Aug}.}
	
	{To meet the aforementioned diverse functionalities, the THz-ISAC system encounters several notable challenges, including but not limited to severe path loss resulting from spreading loss and molecular absorption, limited transmission range, and beam-squint caused by the ultra-wide bandwidth \cite{elbir_thz_jrc_Magazine_Elbir2022Aug, ummimoTareqOverview}. These challenges significantly impact the performance of both S\&C aspects through: 1) {the high path loss leads to extremely low signal-to-noise ratio (SNR) at both radar and communications receivers;} 2) the Doppler shift accompanied by high range sidelobes can trigger false alarms during sensing and introduce inter-carrier interference to the communication systems, and 3) the beam-squint effects cause deviations in the generated beams across different subcarriers, thereby diminishing the array gain and subsequently reducing the spectral efficiency (SE) for communication and the accuracy of direction-finding (DF) for sensing purposes.}

	To address these challenges, the implementation of ISAC concept in massive multiple-input multiple-output (MIMO) systems necessitates large antenna arrays at the base station (BS) to achieve significant beamforming gain \cite{heath2016overview,mimoHeath_spatiallySparse, elbir2022Nov_Beamforming_SPM}. Conversely, massive MIMO systems are also designed with fewer radio frequency (RF) chains to minimize hardware costs. This creates a motivation to develop an efficient ISAC design that carefully balances the system complexity in dealing with challenges such as path loss and beam-squint, while considering the cost implications associated with large arrays. {To reduce this cost, antenna selection is an attractive solution that can select only a high quality subset of antennas to connect the reduced number of RF chains~\cite{elbirQuantizedCNN2019,antennaSelection_MIMO_BAB,antennaSelectionMulticasting_Demir2016Jul,sparseISAC4_Wang2018Aug}. By maintaining a large portion of the same array aperture with fewer antenna elements connected to limited number of RF chains, antenna selection-based systems can achieve a comparable resolution while reducing size, weight, and power-cost (SWAP-C) by utilizing antenna selection diversity~\cite{antennaSelectionHeath_Chen2007Feb}. Nonetheless, removing an element from the antenna array raises the sidelobe levels in the antenna beampattern. This may introduce the ambiguity of resolving target directions in radar systems and the larger interference to other users in communications. Thus, antenna selection in ISAC systems is even more challenging than communications-only and sensing-only systems, and it should take into account the prior information and the constraints related to both S\&C functionalities.
		
		
	}

	To strike a good balance between SE and system complexity, massive MIMO systems employ wideband signal processing, wherein the transceiver architecture is composed of subcarrier-dependent (SD) baseband and subcarrier-independent (SI) analog beamformers. While the SD baseband processing can be carried out on a single hardware platform, the realization of the analog beamformer requires the implementation of phase shifter networks whose size is proportional to the number of subcarriers. As a result, the SI design enjoys a significant hardware simplicity and cost efficiency compared to the SD analog beamformers~\cite{beamSquintAwareHB_SD_You2022Aug}. When the analog beamformers are SI, its design is only constrained by a single (sub-)carrier frequency~\cite{alkhateeb2016frequencySelective,widebandDoAEst_Hybrid_1_Shu2018Feb}. Therefore, the beams generated across the subcarriers point to different directions causing {beam-squint}  phenomenon~\cite{elbir_THZ_CE_ArrayPerturbation_Elbir2022Aug,beamSquint_FeiFei_Wang2019Oct}. The existing techniques to compensate for the impact of beam-squint mostly employ additional hardware components, e.g., time-delayer (TD) networks~\cite{trueTimeDelayBeamSquint,delayPhasePrecoding_THz_Dai2022Mar} and SD phase shifter networks~\cite{beamSquintAwareHB_SD_You2022Aug} to virtually realize SD analog beamformers. 
	However, these approaches are inefficient in terms of cost and power~\cite{delayPhasePrecoding_THz_Dai2022Mar,elbir2023Mar_SPIM_ISAC}. {Beam-squint also occurs in transceivers with antenna selection. Fig.~\ref{fig_AntennaSelectionDiag} illustrates accurate and inaccurate antenna selection configurations for various ISAC scenarios. Although a codebook of selected antennas for different scenarios is available, it should take into account the impact of beam-squint. Otherwise, the squinted beams from the subarray point to different directions may cause inaccurate target detection/estimation for sensing as well as a significant loss in communications rate. }
	
	
	\begin{figure*}[t!]
		\centering
		{\includegraphics[draft=false,width=.6\textwidth]{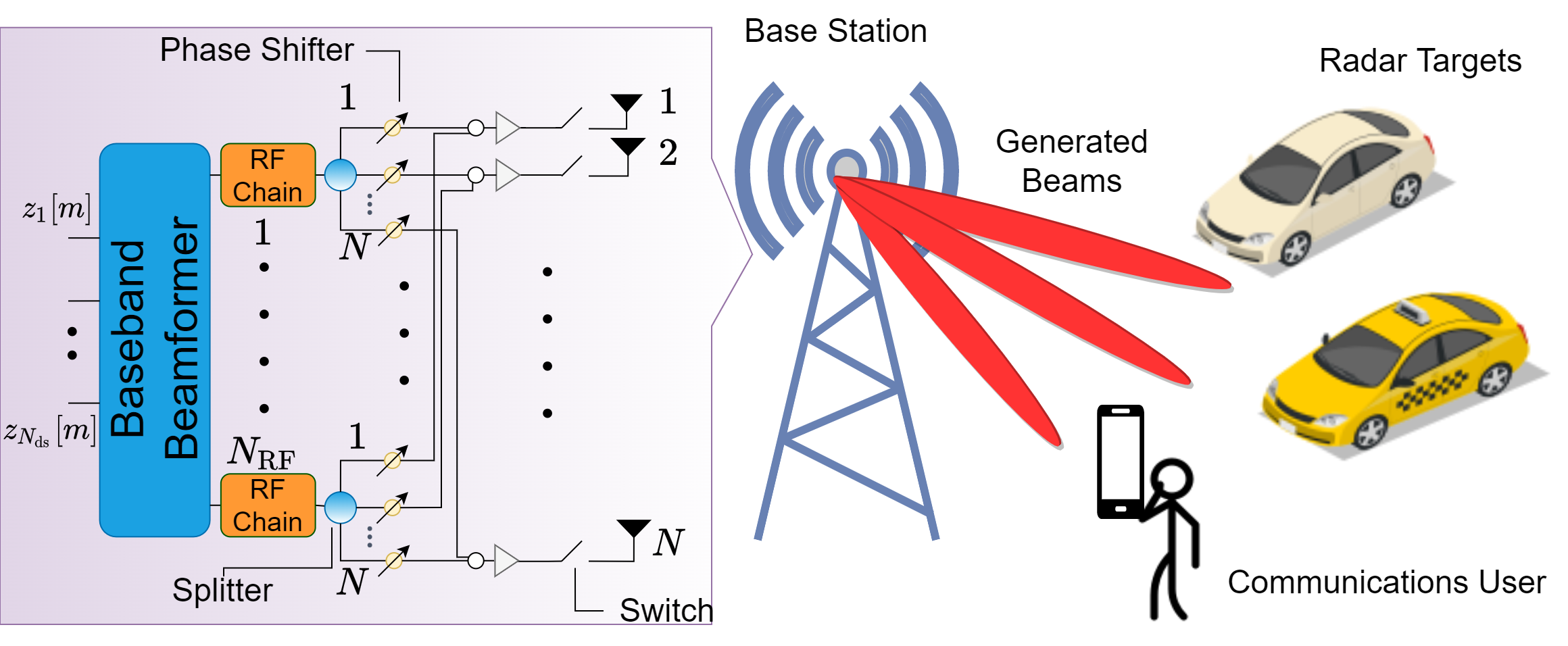}} 
		\caption{ISAC transmitter architecture with hybrid beamforming and antenna selection. 
		}
		\label{fig_BS}
	\end{figure*}

	\subsection{Related Work}
	Recent research includes several works separately on beam-squint compensation~\cite{beamSquintAwareHB_SD_You2022Aug,trueTimeDelayBeamSquint,delayPhasePrecoding_THz_Dai2022Mar,elbir_BSA_OMP_THZ_CE_Elbir2023Feb,elbir_SPIM_MMWAVE_RadarConf_Elbir2022Nov} and sparse array design for communications~\cite{elbirQuantizedCNN2019,sparseISAC4_Wang2018Aug,antennaSelection_MIMO_BAB}, sensing~\cite{elbirIETRSN2019} as well as ISAC~\cite{sparseISAC1_Wang2022Sep,sparseISAC2_Xu2022May,sparseISAC3_Wang2018Dec,sparseISAC4_Wang2018Aug,sparseISAC6_Huang2023Mar}. However, the impact of beam-squint on sparse array design has not been considered in the relevant literature. Because of beam-squint, the performance metric, e.g., SE~\cite{antennaSelection_capacity_Sanayei2007Oct,elbirQuantizedCNN2019,antennaSelection_MIMO_BAB}, channel gain~\cite{antennaSelection_MIMO_Mag_Sanayei2004Oct} and Cram\'er-Rao bound (CRB)~\cite{elbirIETRSN2019,sparseISAC1_Wang2022Sep,sparseISAC2_Xu2022May,antennaSelectionKnapsack}, receive antenna power~\cite{antennaSelection_antennaPowwer_Gao2015Jul}, becomes miscalculated during antenna selection.  Specifically, beam-squint corrupts the array gain and it causes false (deviated) peaks in the spatial domain due to the angular deviations in the generated beams. Therefore, if beam-squint is not compensated properly, the subarray corresponding to the false peaks may differ than that of the optimum subarray.

	Without examining the impact of beam-squint, antenna selection in ISAC scenario is considered in several works~\cite{sparseISAC1_Wang2022Sep,sparseISAC2_Xu2022May,sparseISAC3_Wang2018Dec,sparseISAC4_Wang2018Aug,sparseISAC6_Huang2023Mar}. In particular, antenna selection is performed in~\cite{sparseISAC1_Wang2022Sep} and \cite{sparseISAC2_Xu2022May} via employing the CRB of the target parameter estimation as performance metric. In~\cite{sparseISAC3_Wang2018Dec} and \cite{sparseISAC4_Wang2018Aug}, the authors introduce various ISAC architectures with sparse arrays for single-user configuration. Also, in~\cite{sparseISAC6_Huang2023Mar}, a multi-user, single-target scenario is considered, wherein a fully digital beamforming approach is proposed. Different from the aforementioned model-based techniques, a deep learning (DL)-based approach is proposed in \cite{sparseISAC5_Xu2023Jun}, wherein analog-only beamforming for transmit antenna selection in ISAC scenario is considered. 	Furthermore, most of the antenna selection strategies for ISAC examine whether only analog~\cite{sparseISAC5_Xu2023Jun} or digital~\cite{sparseISAC3_Wang2018Dec,sparseISAC1_Wang2022Sep,sparseISAC4_Wang2018Aug,sparseISAC6_Huang2023Mar} beamformer design without considering hybrid analog/digital beamforming. Although a CRB-based antenna selection and hybrid beamforming is considered in~\cite{sparseISAC2_Xu2022May}, the analog and digital beamformers are not optimized. Besides, DL-based joint hybrid beamforming and antenna selection is studied in various recent works with different settings, e.g., unsupervised learning~\cite{antennaSelection_DL_Unsupervised_Liu2022Jan}, online learning~\cite{antennaSelection_DL_multiUser_Vu2021Jan}, quantized learning model~\cite{elbirQuantizedCNN2019}, and graph learning models~\cite{antennaSelection_DL_Shrestha2023Feb}. However, these works are limited to communications-only scenario and do not consider the impact of beam-squint in wideband systems.

	\subsection{Contributions}
	
	{In this work, we investigate the impact of beam-squint on antenna selection for THz-ISAC hybrid beamforming. The computational complexity of the antenna selection problem is high due to its combinatorial nature, which is addressed by proposed low complexity heuristic solutions that reduce the number of subarray candidates. The performance metric for designing the ISAC hybrid beamformers is the SE of the selected subarray. The main contributions of this work are summarized as follows:
		\begin{itemize}
			\item To design the ISAC hybrid beamformers, we propose a manifold optimization-based approach that incorporates beam-squint compensation (BSC). Unlike recent works on hybrid beamforming with manifold optimization, our approach incorporates wideband processing with BSC. 
			
			\item We devise low complexity algorithms for antenna selection.  In particular, a grouped subarray selection (GSS) approach is proposed, wherein the entire array is divided into distinct, non-overlapping groups, allowing us to select antennas in groups rather than individually, thereby significantly reducing the number of potential subarray configurations. Additionally, we develop a sequential search algorithm to minimize memory requirements during the implementation of antenna selection.
			
			\item By reducing the number of potential subarray configurations, we formulate the antenna selection problem as a classification problem. We develop a learning model with convolutional neural network (CNN) architecture that combines communications and sensing data to efficiently determine the subarray configuration. The CNN model takes the combined communications data (channel matrix) and sensing data (target response vectors) as input. Through training, the CNN model generates the optimal subarray configuration as the output.
			{ 
				\item We examine the impact of beam-squint and subarray configuration in terms of SE of the overall system. In particular, we show, via both theoretical analysis and numerical experiments, that the highest performance can only be achieved if the best subarray is selected and the beam-squint is completely compensated.

			}

		\end{itemize}
	}

	In the remainder of the paper, we present the THz-ISAC architecture with communications and sensing signal model in Section~\ref{sec:SysModel}. Next, we introduce the proposed joint antenna selection and hybrid beamforming approach in Section~\ref{sec:AS_HB}. After presenting various experimental results in Section~\ref{sec:sim},  the paper is finalized with conclusions in Section~\ref{sec:summary}.

	\textit{Notation:} Throughout the paper, we use $(\cdot)^\textsf{T}$, $(\cdot)^{\textsf{H}}$ and $(\cdot)^*$  for transpose and conjugate transpose and complex conjugate operations, respectively. For a matrix $\mathbf{A}$ and vector $\mathbf{a}$; $[\mathbf{A}]_{i,j}$, $[\mathbf{A}]_k$  and $[\mathbf{a}]_l$ correspond to the $(i,j)$-th entry, $k$-th column and $l$-th entry, respectively. Furthermore, $\mathrm{vec}\{\mathbf{A}\}$ denotes the vectorized form of $\mathbf{A}$ with $\mathbf{A} = \mathrm{vec}^{-1}\{\mathrm{vec}\{\mathbf{A}\} \}$. $\mathbb{E}\{\cdot\}$ represent the flooring and expectation operations, respectively. The binomial coefficient is defined as $\footnotesize \left(\begin{array}{c}
	n\\
	k
	\end{array}  \right) = \frac{ n!}{k! (n-k)!}$. An $N\times N$ identity matrix is represented by $\mathbf{I}_{N} $. We denote $\| \cdot \|_0$, $|| \cdot||_2$ and $|| \cdot||_\mathcal{F}$ as the $\ell_0$-norm, $\ell_2$-norm and Frobenious norms, respectively. $\zeta(a) =\frac{\sin N \pi a}{N \sin \pi a} $ is the Drichlet sinc function, and $| \mathbf{A}|$ denotes the determinant of $\mathbf{A}$. $\odot$ and $\otimes$ denote the element-wise Hadamard and Kronecker products, respectively. The Riemannian and Euclidean gradients are represented by  $\nabla_\mathcal{R}$ and $\nabla$, respectively.

	\section{System Model}
	\label{sec:SysModel}
	Consider a wideband ISAC system with hybrid beamforming architecture driven by $N_\mathrm{RF}$ RF chains and $M$ subcarriers as shown in Fig.~\ref{fig_BS}. The BS employs $N$ antennas and aims to simultaneously generate multiple beams towards $T$ targets and a single communications user with $N'$ antennas, for which $N_\mathrm{ds}$ data symbols are transmitted. The BS performs an antenna selection scheme to employ a sparse array of size $K$ out of $N$\footnote{We note here that the selected $K$ antennas are optimized and dedicated to a pair of communications user and radar target for a particular coherence time. The remaining $N-K$ antennas can concurrently be used for another ISAC scenario involving a different user-target pairs available in the network.}\footnote{The number of selected antennas $K$ should satisfy $ T + L \leq N_\mathrm{RF} \leq K \leq N$ to simultaneously generate $T + L$ beams towards $T$ targets and $L$ user path directions.  }. Denoted by $z_i[m]$, $i = 1,\dots, N_\mathrm{ds}$, the transmitted data symbols at the $m$-th subcarrier ($m = \mathcal{M} = \{1,\cdots, M\}$), the BS applies the SD digital beamformer $\mathbf{F}_\mathrm{BB}[m]\in \mathbb{C}^{N_\mathrm{RF}\times N_\mathrm{ds}}$. Using the $K$-element sparse array, the BS applies the SI analog beamformer,  $\mathbf{F}_\mathrm{RF}\in \mathbb{C}^{K\times N_\mathrm{RF}}$ which is realized with fully-connected phase shifter network\footnote{While there are works on partially-connected or subarrayed phase shifter network architectures with~\cite{antennaSelection_MIMO_BAB,elbirQuantizedCNN2019} or without~\cite{fanLiuRadarCommHB_ICASSP2019,elbir2021JointRadarComm} antenna selection, the proposed approach can be easily extended to these architectures via simple modifications in the selection matrix.}. Due to phase-only processing in the phase shifters, the entries of the analog beamformer have the constant modulus property, i.e., $|[\mathbf{F}_\mathrm{RF}]_{i,j}| = 1/\sqrt{K}$ for $i = 1,\cdots, K$ and $j = 1,\cdots, N_\mathrm{RF}$. Then, the $K\times 1$ transmitted signal at the $m$-th subcarrier becomes
	\begin{align}
	\mathbf{g}[m] = \mathbf{F}_\mathrm{RF}\mathbf{F}_\mathrm{BB}[m]\mathbf{z}[m],
	\end{align}
	where $\mathbf{z}[m] = [z_1[m],\cdots, z_{N_\mathrm{ds}}[m]]^\textsf{T}$, and $\mathbb{E}\{\mathbf{z}[m]\mathbf{z}^\textsf{H}[m]\} = \frac{1}{N_\mathrm{ds}}\mathbf{I}_{N_\mathrm{ds}}$.
	
	\subsection{Communications Model}
	Denote the downlink THz channel matrix between the BS and the communications user as $\overline{\mathbf{H}}[m]\in \mathbb{C}^{N'\times N}$. Then, the channel matrix with selected antennas is
	\begin{align}
	\mathbf{H}[m] = \overline{\mathbf{H}}[m] \mathbf{Q},
	\end{align}
	where  $\mathbf{Q}\in \{0,1\}^{N\times K}$ is the selection matrix. Specifically, for the $(n,k)$-th element of $\mathbf{Q}$,  $[\mathbf{Q}]_{n,k} = 1$ represents that the $n$-th transmit antenna is the $k$-th selected  antenna for $n\in \{1,\cdots, N\}$ and $k \in \{1, \cdots, K\}$. Then, the $N'\times  1$ received signal vector at the user becomes
	\begin{align}
	\mathbf{y}[m] &= \overline{\mathbf{H}}[m]\mathbf{Q}\mathbf{g}[m] + \mathbf{n}[m] \nonumber \\
	& = \mathbf{H}[m]\mathbf{F}_\mathrm{RF}\mathbf{F}_\mathrm{BB}[m]\mathbf{z}[m] + \mathbf{n}[m],
	\end{align}
	where $\mathbf{n}[m]\in \mathbb{C}^{N'}$ denotes the temporally and spatially white additive zero-mean Gaussian noise vector with variance $\sigma^2$.

	\subsubsection{THz Channel}
	Channel modeling in THz band has been a challenging task largely because of the lack of realistic measurement campaigns~\cite{thz_Akyildiz2022May,thz_channel_Rappaport3_Ju2021Jun,thz_channel_Rappaport1_Ju2022May}. 
	In  \cite{thz_channel_Rappaport1_Ju2022May,thz_channel_Rappaport3_Ju2021Jun},  it is shown that a single dominant line-of-sight (LoS) path with a few non-LoS (NLoS) multipath components survive at the receiver in outdoor scenarios for sub-THz frequencies~\cite{ummimoHBThzSVModel}. In a general scenario, e.g., indoor, multipath channels can also arise, the gains of  LoS and NLoS paths are comparable~\cite{ummimoTareqOverview,thz_mmWave_path_Comparison_Yan2020Jun,elbir2023Apr_aUnified_BS_THz_HB}. Therefore, we assume a general scenario, wherein the THz channel matrix $\overline{\mathbf{H}}[m]$ includes the contribution of $L$ multipath scatterers as
	\begin{align}
	\label{channel1}
	\overline{\mathbf{H}}[m] = \sqrt{\frac{N'N}{L}} \sum_{l = 1}^{L} \alpha_{l,m} \mathbf{a}'(\phi_l)\mathbf{a}^\textsf{H}(\theta_l),
	\end{align}
	where $\alpha_{l,m}\in \mathbb{C}$ denotes the gain of the $l$-th path, and it can be defined for a LoS path as 
	\begin{align}
	\alpha_{l,m}^{\mathrm{LoS}} = 
	\frac{c}{4 \pi f_m \bar{d}_l }  e^{- \frac{1}{2}k_\mathrm{abs} (f_m) \bar{d}_l  }  e^{- \mathrm{j}\frac{2\pi f_m}{c}\bar{d}_l  } 
	\end{align} 
	where $c$ is the speed-of-light, $k_\mathrm{abs}(f_m)$ is the SD molecular absorption coefficient for the $m$-th subcarrier frequency $f_m$ and $\bar{d}_l$ is the transmission distance~\cite{thz_Akyildiz2022May,thz_channel_Rappaport1_Ju2022May}. For NLoS paths, the expected path gain is given by
	\begin{align}
	\mathbb{E} \{|\alpha_{l,m}^{\mathrm{NLoS}}|^2\} = \left(\frac{c}{4 \pi f_m \bar{d}_l  }\right)^2 e^{-k_\mathrm{abs} (f_m) \bar{d}_l  }  e^{- \frac{\bar{\tau}_l}{\bar{\Gamma}}  },
	\end{align}
	where $\bar{\tau}$  is the time of arrival of the $l$-th path while  $\bar{\Gamma}$  denotes the ray decay factor~\cite{ummimoTareqOverview}. We note here that the proposed hybrid beamforming techniques for THz channel are also applicable for both narrowband and wideband mmWave systems. 
	
	In (\ref{channel1}), $\mathbf{a}'(\phi_l)\in \mathbb{C}^{N'}$ and $\mathbf{a}(\theta_l)\in \mathbb{C}^{N}$ are the steering vectors corresponding to the physical direction-of-arrival (DoA) ($\phi_l$) and direction-of-departure (DoD) angles ($\theta_l$) of the $l$-th paths, respectively. The $n$-th element of $\mathbf{a}(\theta_l)$ for a uniform linear array (ULA) is given by
	\begin{align}
	[\mathbf{a}(\theta_l)]_n = \frac{1}{\sqrt{N}} \exp \{-\mathrm{j} 2\pi \frac{d}{\lambda_c}(n-1)\sin \theta_l \}, \label{steeringVec1}
	\end{align}
	where $n = 1,\cdots, N$,  $\lambda_c$ is the wavelength of the central subcarrier frequency, i.e., $\lambda_c = \frac{c}{f_c}$, where $f_c$ is the carrier frequency and $d$ is the antenna element spacing, which is typically selected as $d = \lambda_c/2$. Note that the receive steering vector $\mathbf{a}'(\phi_l)$ can be defined similarly.

	\subsubsection{Beam-Squint Effect}
	\label{sec:beamSplit} 
	In wideband transmission, it is typically assumed that a common analog beamformer is designed corresponding to a single wavelength for all subcarriers, i.e., $\lambda_1 = \cdots =  \lambda_M = \frac{c}{f_c}$. However, this assumption no longer holds when bandwidth is so large that the beams generated at different subcarriers squint and they point to different directions in spatial domain~\cite{beamSquint_FeiFei_Wang2019Oct,elbir_thz_jrc_Magazine_Elbir2022Aug,delayPhasePrecoding_THz_Dai2022Mar,elbir_THZ_CE_ArrayPerturbation_Elbir2022Aug}. If a similar beamforming architecture, employing SI analog beamformer and SD digital beamformers, is also utilized by the user, the same beam-squint effect is also observed at the user. The amount of beam-squint in the spatial domain is SD and it becomes larger as $|f_m-f_c|$ increases. Thus, we define the SD beam-squinted DoA and DoD angles in spatial domain as $\sin\varphi_{l,m}$ and $\sin \vartheta_{l,m}$, respectively. Then, the relationship between the spatial and physical directions ($\sin\phi_l, \sin\theta_l$) is given as
	\begin{align}
	\sin\varphi_{l,m} =\eta_m \sin\phi_l, \hspace{10pt}
	\sin\vartheta_{l,m} = \eta_m\sin\theta_l, \label{beamSplitforAngles}
	\end{align}
	where $\eta_m = \frac{f_m}{f_c}$, $f_m = f_c + \frac{B}{M}(m - 1 - \frac{M-1}{2})$ is the $m$-th subcarrier frequency for the system bandwidth $B$.  We can see beam-squint is mitigated if the spatial $(\sin\varphi_{l,m},\sin\vartheta_{l,m})$ and physical directions $(\sin\phi_l, \sin\theta_l)$ are equal, i.e., $\eta_m = 1$. 
	
	
	Under the effect of beam-squint, the $n$-th entry of the SD steering vector $\mathbf{a}(\vartheta_{l,m})$ is given by
	\begin{align}
	[\mathbf{a}(\vartheta_{l,m})]_n &= \frac{1}{\sqrt{N}} \exp \left\{- \mathrm{j} \frac{2\pi d}{\lambda_m } (n-1) \sin\theta_l\right \} 
	\nonumber\\	& 
	=\frac{1}{\sqrt{N}} \exp\left\{- \mathrm{j}\pi  \frac{ f_m}{f_c }(n-1)\sin \theta_l \right\} \nonumber \\
	&=\frac{1}{\sqrt{N}} \exp\left\{- \mathrm{j}\pi (n-1)\eta_m \sin\theta_l \right\},\label{steringVec_aT}
	\end{align}
	where  $\lambda_m = \frac{c}{f_m}$ is the wavelength of the $m$-th subcarrier. Comparing (\ref{steeringVec1}) and (\ref{steringVec_aT}) yields that the deviation in the spatial directions due to beam-squint can be compensated by 
	exploiting the phase terms of the steering vectors as will be discussed in Sec.~\ref{sec:BeamSplitMitigation}.

	For communications-only problem, the hybrid beamforming design  aims to maximize the SE which is defined for the $m$-th subcarrier as 
	\begin{align}
	\mathrm{SE}[m] &= \log_2 \big|\mathbf{I}_{N'} + \frac{1}{N_\mathrm{ds} \sigma^2} \mathbf{H}[m] \mathbf{F}_\mathrm{RF}\mathbf{F}_\mathrm{BB}[m] \nonumber \\
	&\hspace{40pt} \times \mathbf{F}_\mathrm{BB}^\textsf{H}[m]{\mathbf{F}_\mathrm{RF}}^\textsf{H} {\mathbf{H}}^\textsf{H}[m]       \big|, \label{SE}
	\end{align}
	for which the SE of the overall system is $\mathrm{SE} = \sum_{m =1}^{M}\mathrm{SE}[m]$.  We note that maximizing the SE can be achieved by exploiting the similarity between the hybrid beamformer $\mathbf{F}_\mathrm{RF}\mathbf{F}_\mathrm{BB}[m]$ and the unconstrained communications-only beamformer $\mathbf{F}_\mathrm{C}[m]\in \mathbb{C}^{K\times N_\mathrm{ds}}$~\cite{mimoHeath_spatiallySparse,heath2016overview,elbir2022Nov_Beamforming_SPM}. In particular, $\mathbf{F}_\mathrm{C}[m]$ is the subarray beamformer as
	\begin{align}
	\mathbf{F}_\mathrm{C}[m] =\mathbf{Q}^\textsf{T} \overline{\mathbf{F}}_\mathrm{C}[m],
	\end{align} where $\overline{\mathbf{F}}_\mathrm{C}[m]\in \mathbb{C}^{N \times N_\mathrm{ds}}$ denotes the communications-only beamformer corresponding to the full array, and it can be directly obtained from the right singular matrix of $\overline{\mathbf{H}}[m]$ via singular value decomposition (SVD)~\cite{heath2016overview}.
	
	{\textit{Assumption 1:} We assume that the THz channel matrix of the full array, i.e., $\overline{\mathbf{H}}[m]$ is available for ISAC beamformer design. This can be achieved either model-based techniques~\cite{dovelos_THz_CE_channelEstThz2,thz_ce_beamSquint_Tan2021May,elbir_BSA_OMP_THZ_CE_Elbir2023Feb} or learning-based approaches~\cite{elbir_THZ_CE_ArrayPerturbation_Elbir2022Aug,thz_CE_GAN_Balevi2021Jan}. It is also worth noting that the complete channel matrix can be constructed by cycling the $N_\mathrm{RF}$ RF chains among $N$ antennas during channel training. In other words, the RF chains are first connected to the first $N_\mathrm{RF}$ antennas during the first part of the training sequence, then the second $N_\mathrm{RF}$ antennas, and so on~\cite{antennaSelection_Capacity_Molisch2005Jul,heath2016overview}.
		
	}
	
	\subsection{Sensing Model}
	While communicating with the user, the ISAC system aims to deliver as high SNR as possible to targets for sensing~\cite{mimoRadar_WidebandYu2019May,elbir2021JointRadarComm}. To that end, the BS transmits probing signals to sense the targets in the environment. Let $\tilde{\mathbf{X}}_t[m]\in \mathbb{C}^{N\times T_\mathrm{S}}$ be the transmitted sensing signal, where $T_\mathrm{S}$ is the number of snapshots. Then, the $K\times T_\mathrm{S}$ received echo signal from $T$ targets is
	\begin{align}
	\tilde{\mathbf{Y}}[m] = \sum_{t = 1}^{T} \beta_t \mathbf{a}(\Phi_t) \mathbf{a}^\textsf{T}(\Phi_t) \tilde{\mathbf{X}}[m] + \tilde{\mathbf{N}}[m],\label{receivedSignalSensing}
	\end{align}
	where $\beta_t$ represents the reflection coefficient, i.e., radar cross section, of the target, $\mathbf{a}(\Phi_t)\in \mathbb{C}^{N}$ denotes the steering vector corresponding to the $t$-th target at the direction $\Phi_t$ and $\tilde{\mathbf{N}}[m]\in \mathbb{C}^{N\times T_\mathrm{S}}$ denotes  the additive noise term. The estimation of the target directions can be performed via both model-based~\cite{music,mimoRadar_WidebandYu2019May,millidegreDOA_EST_Chen2021Aug} and model-free learning-based~\cite{elbir_DL_MUSIC} techniques available in the literature. Once th target directions are estimated as $\{\hat{\Phi}_t\}_{t \in \mathcal{T}} $, where $\mathcal{T} = \{1,\cdots, T\}$, then the sensing-only  beamformer $\overline{\mathbf{F}}_\mathrm{S}\in \mathbb{C}^{N\times T}$ is constructed as 
	\begin{align}
	\overline{\mathbf{F}}_\mathrm{S} = \left[\mathbf{a}(\hat{\Phi}_1),\cdots, \mathbf{a}(\hat{\Phi}_T)  \right].
	\end{align}
	Then, the sensing-only subarray beamformer $	\mathbf{F}_\mathrm{S}\in \mathbb{C}^{K\times T}$ can be given by
	\begin{align}
	\mathbf{F}_\mathrm{S} = \mathbf{Q}^\textsf{T}\overline{\mathbf{F}}_\mathrm{S}.
	\end{align}

	{\textit{Assumption 2:} We  assume that  the sensing-only full array beamformer $\overline{\mathbf{F}}_\mathrm{S}$ is available. That is to say, the target directions are acquired during the search operation of the radar prior to the beamformer design.  {Although the relevant literature on the direction estimation is mostly limited to beam-squint-free scenario~\cite{millidegreDOA_EST_Chen2021Aug,music,angleDomain_doaEst_AngleDomain_Fan2017Dec},  a beam-squint-aware multiple signal classification (BSA-MUSIC) technique is introduced recently in~\cite{elbir2023Mar_SPIM_ISAC}  for the compensation of beam-squint in direction estimation problem. }
		
	}

	\subsection{Problem Formulation}
	Our aim in this work is to jointly optimize a subarray at the BS as well as designing hybrid beamformers, which can be achieved via maximizing the SE of the overall system. By exploiting the similarity between the hybrid beamformer $\mathbf{F}_\mathrm{RF}\mathbf{F}_\mathrm{BB}[m]$ and $\mathbf{F}_\mathrm{C}[m]$, $\mathbf{F}_\mathrm{S}$, the joint antenna selection and hybrid beamforming problem is written as 
	\begin{subequations}
		\begin{align}
		\minimize_{\mathbf{Q}, \mathbf{F}_\mathrm{RF}, \{\mathbf{F}_\mathrm{BB}[m], \mathbf{D}[m]\}_{m\in \mathcal{M}}  } & \sum_{m = 1}^{M} \bigg(  \varepsilon\|\mathbf{F}_\mathrm{RF}\mathbf{F}_\mathrm{BB}[m] - \mathbf{F}_\mathrm{C}[m]    \|_\mathcal{F}  \nonumber \\
		& \hspace{-70pt} + (1 - \varepsilon) \|   \mathbf{F}_\mathrm{RF}\mathbf{F}_\mathrm{BB}[m] - \mathbf{F}_\mathrm{S}\mathbf{D}[m] \|_\mathcal{F}   \bigg) \nonumber \\
		& \hspace{-75pt} \subjectto  \sum_{m = 1}^{M}\|\mathbf{F}_\mathrm{RF}\mathbf{F}_\mathrm{BB}[m]     \|_\mathcal{F} = M N_\mathrm{ds}, \label{c_power}  \\
		& \hspace{-20pt} |[\mathbf{F}_\mathrm{RF}]_{i,j}| = 1/\sqrt{N}, \label{c_constantMod} \\
		& \hspace{-20pt} \mathbf{D}[m] \mathbf{D}^\textsf{H}[m] = \mathbf{I}_T,	\label{unity} \\
		& \hspace{-20pt} \mathbf{F}_\mathrm{C}[m] =\mathbf{Q}^\textsf{T} \overline{\mathbf{F}}_\mathrm{C}[m],	 \\
		& \hspace{-20pt} 	\mathbf{F}_\mathrm{S} =\mathbf{Q}^\textsf{T} \overline{\mathbf{F}}_\mathrm{S}, \\
		& \hspace{-20pt} [\mathbf{Q}]_{n,k}\in \{0,1\} , \label{c_combinatorial} \\
		& \hspace{-20pt} \| \mathrm{vec}\{\mathbf{Q} \}  \|_0 = K,  \label{c_combinatorial2}
		\end{align}
		\label{opt1}
	\end{subequations}
	\noindent where $\mathbf{D}[m]\in \mathbb{C}^{T\times N_\mathrm{ds}}$ is a unitary matrix (i.e, $\mathbf{D}[m] \mathbf{D}^\textsf{H}[m] = \mathbf{I}_T$) and it provides the change of dimensions between $\mathbf{F}_\mathrm{S}$ and $\mathbf{F}_\mathrm{C}[m]$. In (\ref{opt1}), $0\leq \varepsilon \leq 1$ is the trade-off parameter between communications and sensing tasks. Specifically,  $\varepsilon =1$  ($\varepsilon = 0$) corresponds to communications-only (sensing-only) design.  The procedure of determining $\varepsilon$  includes the ratio of allocated resources, such as power~\cite{tradeoff_parameterSelection_Chiriyath2015Sep} and  signal durations of the coherent processing intervals~\cite{tradeoff_CPI_Dokhanchi2019Feb}.
	
	The problem in (\ref{opt1}) falls to the class of mixed-integer non-convex programming (MINCP){, which is  difficult to solve due to several matrix variables $\mathbf{Q}, \mathbf{F}_\mathrm{RF}, \mathbf{F}_\mathrm{BB}[m], \mathbf{D}[m]$ ans non-convex constraints. In particular, the constant-modulus constraints for the  analog beamformers $\mathbf{F}_\mathrm{RF}$ in (\ref{c_constantMod}) indicate that the amplitude of the analog beamformer weights are a constant. Furthermore, the antenna selection matrix $\mathbf{Q}$ has binary values as in (\ref{c_combinatorial}) and, the number of non-zero terms in $\mathbf{Q}$ equals to $K$ as in  (\ref{c_combinatorial2}). Following these considerations, we introduce an effective and computationally-efficient solution in the remainder of this work.  } 
	
	
	\section{Joint Antenna Selection and  \\ Beamformer Design}
	\label{sec:AS_HB}
	In order to provide an effective solution, we first divide the problem in (\ref{opt1}) into two subproblems, hybrid beamforming and antenna selection. In hybrid beamforming design, we first formulate the subproblem as a manifold optimization problem for a given subarray configuration, wherein the hybrid analog/digital beamformers are alternatingly optimized as well as the impact of beam-squint is compensated.
	
	In antenna selection, we are interested in selecting $K$  out of $N$ antenna elements at the BS. This yields $P = \footnotesize \left(\begin{array}{c}
	N \\
	K
	\end{array}  \right) = \frac{N!}{K! (N - K)!  }$ possible subarray configurations. Therefore, antenna selection problem can be viewed as a classification problem with $P$ classes. Define $\mathcal{Q} = \{\mathbf{Q}_1,\cdots, \mathbf{Q}_P  \}$ as the set of all possible subarray configurations, where $ \mathbf{Q}_p$ represents the selection matrix $\mathbf{Q}$ for the $p$-th configuration as $\mathbf{Q}_p = [\mathbf{q}_1^p,\cdots, \mathbf{q}_K^p ] $. Here,  $\mathbf{q}_k^p = [0, \cdots, q_{n,k}^p,\cdots, 0]^\textsf{T}$, and we have $q_{n,k}^p = 1$ corresponding to the $k$-th element of the subarray as the selected $n$-th transmit antenna for the $p$-th configuration.


	\subsection{Hybrid Beamformer Design}
	We first define the hybrid beamformers for the $p$-th subarray configuration as $\mathbf{F}_\mathrm{RF}^{(p)}\in \mathbb{C}^{K\times N_\mathrm{RF}}$ and $\mathbf{F}_\mathrm{BB}^{(p)}[m]\in \mathbb{C}^{N_\mathrm{RF}\times N_\mathrm{ds}}$. Next, we define the cost function in (\ref{opt1}) for the $p$-th configuration and the $m$-th subcarrier as 
	\begin{align}
	f(p,m) &=  \varepsilon \| \mathbf{F}_\mathrm{RF}^{(p)} \mathbf{F}_\mathrm{BB}^{(p)}[m] - \mathbf{F}_\mathrm{C}^{(p)} [m]   \|_\mathcal{F} \nonumber \\
	& \hspace{15pt}+ (1- \varepsilon) \| \mathbf{F}_\mathrm{RF}^{(p)} \mathbf{F}_\mathrm{BB}^{(p)}[m] - \mathbf{F}_\mathrm{S}^{(p)} \mathbf{D}^{(p)}[m]   \|_\mathcal{F},
	\end{align}
	where $\mathbf{F}_\mathrm{C}^{(p)} [m]  =\mathbf{Q}_p \overline{\mathbf{F}}_\mathrm{C}[m]$ and  $\mathbf{F}_\mathrm{S}^{(p)}  =\mathbf{Q}_p \overline{\mathbf{F}}_\mathrm{S}$. Using triangle inequality, the following can be obtained, i.e., 
	\begin{align}
	&f(p,m) \geq  \| \varepsilon \mathbf{F}_\mathrm{RF}^{(p)} \mathbf{F}_\mathrm{BB}^{(p)}[m] - \varepsilon\mathbf{F}_\mathrm{C}^{(p)} [m] \nonumber \\
	& + (1-\varepsilon) \mathbf{F}_\mathrm{RF}^{(p)} \mathbf{F}_\mathrm{BB}^{(p)}[m] -  (1-\varepsilon) \mathbf{F}_\mathrm{S}^{(p)} \mathbf{D}^{(p)}[m]   \|_\mathcal{F} \nonumber \\
	&=   \|  \mathbf{F}_\mathrm{RF}^{(p)} \mathbf{F}_\mathrm{BB}^{(p)}[m] - \underbrace{\varepsilon\mathbf{F}_\mathrm{C}^{(p)} [m]  -  (1-\varepsilon) \mathbf{F}_\mathrm{S}^{(p)} \mathbf{D}^{(p)}[m] }_{\triangleq{\mathbf{F}}_\mathrm{SC}^{(p)}[m] }  \|_\mathcal{F}, \label{fCost1}
	\end{align}
	where we define ${\mathbf{F}}_\mathrm{SC}^{(p)}[m] \in \mathbb{C}^{K\times N_\mathrm{ds}}$ as the  joint sensing-communications (JSC) beamformer, which involves the combination of ${\mathbf{F}}_\mathrm{C}^{(p)}[m]$ and ${\mathbf{F}}_\mathrm{S}^{(p)}$~\cite{elbir2021JointRadarComm,jrc_TCOM_Liu2020Feb,fanLiuRadarCommHB_ICASSP2019} as
	\begin{align}
	{\mathbf{F}}_\mathrm{SC}^{(p)}[m] = \varepsilon{\mathbf{F}}_\mathrm{C}^{(p)}[m] + (1-\varepsilon) {\mathbf{F}}_\mathrm{S}^{(p)}\mathbf{D}^{(p)}[m]. \label{FSC}
	\end{align}
	
	Since we have $f(p,m) \geq  \|  \mathbf{F}_\mathrm{RF}^{(p)} \mathbf{F}_\mathrm{BB}^{(p)}[m] - {\mathbf{F}}_\mathrm{SC}^{(p)}[m]   \|_\mathcal{F} $ from (\ref{fCost1}), maximizing $\|  \mathbf{F}_\mathrm{RF}^{(p)} \mathbf{F}_\mathrm{BB}^{(p)}[m] - {\mathbf{F}}_\mathrm{SC}^{(p)}[m]   \|_\mathcal{F} $ is equivalent to maximizing $f(p,m)$. Thus,  we can write the hybrid beamforming design problem for the $p$-th subarray configuration as 
	\begin{align}
	\minimize_{\mathbf{F}_\mathrm{RF}^{(p)}, \mathbf{F}_\mathrm{BB}^{(p)}[m], \mathbf{D}^{(p)}[m] } &\; \sum_{m  = 1}^{M}||\mathbf{F}_\mathrm{RF}^{(p)}\mathbf{F}_\mathrm{BB}^{(p)}[m] - {\mathbf{F}}_\mathrm{SC}^{(p)}[m]     \|_\mathcal{F} \nonumber \\
	\subjectto & |[\mathbf{F}_\mathrm{RF}^{(p)}]_{k,n}| = 1/\sqrt{K}, \nonumber \\
	& \sum_{m = 1}^M \|\mathbf{F}_\mathrm{RF}^{(p)}\mathbf{F}_\mathrm{BB}^{(p)}[m]\|_\mathcal{F} = M N_\mathrm{ds},\nonumber \\
	& \mathbf{D}^{(p)}[m]{\mathbf{D}^{(p)}}^\textsf{H}[m] = \mathbf{I}_T, \label{optHB}
	\end{align}
	which can be written in a compact form as 
	\begin{align}
	\minimize_{\mathbf{F}_\mathrm{RF}^{(p)}, \widetilde{\mathbf{F}}_\mathrm{BB}^{(p)}, \widetilde{\mathbf{D}}^{(p)} } &\;||\mathbf{F}_\mathrm{RF}^{(p)}\widetilde{\mathbf{F}}_\mathrm{BB}^{(p)}- \widetilde{\mathbf{F}}_\mathrm{SC}^{(p)}    \|_\mathcal{F} \nonumber \\
	\subjectto & |[\mathbf{F}_\mathrm{RF}^{(p)}]_{k,n}| = 1/\sqrt{K}, \nonumber \\
	& \|\mathbf{F}_\mathrm{RF}^{(p)}\widetilde{\mathbf{F}}^{(p)}\|_\mathcal{F} = M N_\mathrm{ds},\nonumber \\
	& \widetilde{\mathbf{D}}^{(p)}\widetilde{\mathbf{D}}^{(p)^\textsf{H}} = \mathbf{I}_{T}, \label{optHB2}
	\end{align}
	where $\widetilde{\mathbf{F}}_\mathrm{BB}^{(p)} = \left[\mathbf{F}_\mathrm{BB}^{(p)}[1], \cdots, \mathbf{F}_\mathrm{BB}^{(p)}[M] \right]$, $\widetilde{\mathbf{F}}_\mathrm{SC}^{(p)} = \left[\mathbf{F}_\mathrm{SC}^{(p)}[1], \cdots, \mathbf{F}_\mathrm{SC}^{(p)}[M] \right]$ and $\widetilde{\mathbf{D}}^{(p)} = \left[\mathbf{D}^{(p)}[1], \cdots, \mathbf{D}^{(p)}[M] \right]$	are $K\times MN_\mathrm{ds}$, $K\times MN_\mathrm{ds}$ and $T\times MN_\mathrm{ds}$ matrices, respectively.
	
	Now, the optimization problem in (\ref{optHB2}) can be solved effectively via manifold optimization techniques~\cite{manopt,hybridBFAltMin,elbir2022Nov_Beamforming_SPM}. To solve (\ref{optHB2}), we follow an alternating  technique, wherein the unknown variables $\mathbf{F}_\mathrm{RF}^{p}$, $\widetilde{\mathbf{F}}_\mathrm{BB}^{(p)}$ and $\widetilde{\mathbf{D}}^{(p)}$ are estimated one by one while the remaining terms are fixed. 
	
	\subsubsection{Solve for $\mathbf{F}_\mathrm{RF}^{(p)}$}
	In order to solve for ${\mathbf{F}}_\mathrm{RF}^{(p)}$ via manifold optimization, we first define $\mathbf{f}^{(p)} = \mathrm{vec}\{ \mathbf{F}_\mathrm{RF}^{(p)} \}\in \mathbb{C}^{KN_\mathrm{RF}}$ as the vectorized form of $\mathbf{F}_\mathrm{RF}^{(p)}$. By exploiting the unit-modulus constraint of the analog beamformer, the search space for $\mathbf{f}^{(p)}$ is regarded as \textit{Riemannian submanifold} $\mathcal{R}$ of the complex plane $\mathbb{C}^{K N_\mathrm{RF}}$ as~\cite{manopt,manopt2_Zhou2017Apr}
	\begin{align}
	\mathcal{R} = \{\mathbf{f}^{(p)}\in \mathbb{C}^{KN_\mathrm{RF}} : | [\mathbf{f}^{(p)}]_1|=\cdots = |[\mathbf{f}^{(p)}]_{KN_\mathrm{RF}}| = \frac{1}{\sqrt{K}}  \}.
	\end{align}  Then, by following a conjugate gradient descent technique~\cite{hybridBFAltMin,manopt3_multiUser_Vandendorpe_Du2019Jul}, $\mathbf{f}^{(p)}$ can be optimized iteratively, and in the $i$-th iteration, we have 
	\begin{align}
	\mathbf{f}_i^{(p)} = \frac{\mathbf{f}_i^{(p)} + \dot{a}_i \boldsymbol{\xi}_i (\mathbf{f}_i, \widetilde{\mathbf{F}}_\mathrm{BB}^{(p)}, \widetilde{\mathbf{F}}_\mathrm{SC}^{(p)})  }{|\mathbf{f}_i^{(p)} - \dot{a}_i \mathbf{f}_i |}, \label{f_i_MO_update}
	\end{align}
	where $\dot{a}_i$ is Armijo backtracking line search step size~\cite{armijoStepSize_Truong2021Dec} and $\boldsymbol{\xi}_i (\mathbf{f}_i^{(p)}, \widetilde{\mathbf{F}}_\mathrm{BB}^{(p)}, \widetilde{\mathbf{F}}_\mathrm{SC}^{(p)}) \in \mathbb{C}^{KN_\mathrm{RF}}$ is the directional gradient vector~\cite{manopt}, and it depends on the Riemannian gradient of $\mathbf{f}_i^{(p)}$, i.e.,
	$\nabla_\mathcal{R} \mathbf{f}_i^{(p)}$, which is defined as
	\begin{align}
	\nabla_\mathcal{R} \mathbf{f}_i^{(p)} = \nabla \mathbf{f}_i^{(p)} - \operatorname{Re}\{\nabla \mathbf{f}_i^{(p)} \odot \mathbf{f}_i^{(p)^*}  \} \odot \mathbf{f}_i^{(p)}, \label{gradR}
	\end{align}
	where $\nabla\mathbf{f}_i^{(p)}$ denotes the Euclidean gradient of $\mathbf{f}_i$ as
	\begin{align}
	\nabla\mathbf{f}_i^{(p)} = -2 \mathbf{B}^{(p)}   \left( \mathbf{f}_\mathrm{SC}^{(p)} - \mathbf{B}^{(p)}  \mathbf{f}_i^{(p)}   \right), \label{conj_gradient}
	\end{align}
	where $\mathbf{B}^{(p)}   = \widetilde{\mathbf{F}}_\mathrm{BB}^{(p)^\textsf{T}} \otimes \mathbf{I}_{K} \in \mathbb{C}^{KMN_\mathrm{ds}\times KN_\mathrm{RF}}$ and $\mathbf{f}_\mathrm{SC}^{(p)} = \mathrm{vec}\{\widetilde{\mathbf{F}}_\mathrm{SC}^{(p)} \}\in \mathbb{C}^{KMN_\mathrm{ds}}$. 
	
	Given $\widetilde{\mathbf{F}}_\mathrm{BB}^{(p)}$ and $\widetilde{\mathbf{F}}_\mathrm{SC}^{(p)}$, the optimization process can be initialized for $i = 0$ by selecting $\mathbf{f}_0^{(p)}$ as $[\mathbf{f}_0^{(p)}]_k = e^{j\Psi_k}$, where $\Psi_k \sim \mathrm{uniform}([0,2\pi))$ for $k = 1,\cdots, KN_\mathrm{RF}$. The complexity of computing (\ref{f_i_MO_update}) is mainly due to the computation of the conjugate gradient in (\ref{conj_gradient}). Therefore, the computational complexity order of optimizing $\mathbf{F}_\mathrm{RF}^{(p)}$ is $O(N_\mathrm{iter}^A K^2 N_\mathrm{RF} M N_\mathrm{ds}) $, where $N_\mathrm{iter}^A$ is the number of iterations~\cite{elbir2021JointRadarComm,manopt3_multiUser_Vandendorpe_Du2019Jul}.

	\subsubsection{Solve for $\widetilde{\mathbf{F}}_\mathrm{BB}^{(p)}$ and Beam-Squint Compensation (BSC)}
	\label{sec:BeamSplitMitigation}
	Since the analog beamformer $\mathbf{F}_\mathrm{RF}^{(p)}$ is SI, beam-squint occurs. In order to compensate for the impact of beam-squint, we design the baseband beamformer $\mathbf{F}_\mathrm{BB}^{(p)}[m]$ accordingly such that the impact of beam-squint in the analog domain is conveyed to the baseband which is SD.		To this end, we first obtain $\mathbf{F}_\mathrm{BB}^{(p)}[m]$ from  ${\mathbf{F}}_\mathrm{SC}^{(p)}[m]$ and $\mathbf{F}_\mathrm{RF}^{(p)}$. Then, update the baseband beamformer by utilizing SD analog beamformer, which can be virtually computed from the SI analog beamformer $\mathbf{F}_\mathrm{RF}^{(p)}$~\cite{elbir2023Mar_SPIM_ISAC}.

	Given $\mathbf{F}_\mathrm{BB}^{(p)}[m]$ from  $\widetilde{\mathbf{F}}_\mathrm{SC}^{(p)}$ and $\mathbf{F}_\mathrm{RF}^{(p)}$, a straightforward solution for ${\mathbf{F}}_\mathrm{BB}^{(p)}[m]$ is
	\begin{align}
	\breve{\mathbf{F}}_\mathrm{BB}^{(p)}[m] = \left( \mathbf{F}_\mathrm{RF}^{(p)} \right)^\dagger  {\mathbf{F}}_\mathrm{SC}^{(p)}[m]. \label{fbb1}
	\end{align}
	Now, we define the SD analog beamformer as $\breve{\mathbf{F}}_\mathrm{RF}^{(p)}[m]\in \mathbb{C}^{K\times N_\mathrm{RF}}$, which can be found from $\mathbf{F}_\mathrm{RF}^{(p)}$ as
	\begin{align}
	\breve{\mathbf{F}}_\mathrm{RF}^{(p)}[m] = \frac{1}{\sqrt{K}} \boldsymbol{\Omega}_m^{(p)},\label{frf2}
	\end{align}
	where $\boldsymbol{\Omega}_m^{(p)}\in \mathbb{C}^{K\times N_\mathrm{RF}}$ includes the phase information of the SI beamformer $\mathbf{F}_\mathrm{RF}^{(p)}$ as 
	\begin{align}
	[\boldsymbol{\Omega}_m^{(p)}]_{k,n} = \exp \{\mathrm{j} \eta_m \angle \{[\mathbf{F}_\mathrm{RF}^{(p)}]_{k,n} \}  \},\label{frf3}
	\end{align}
	for $k = 1,\cdots, K$ and $n = 1,\cdots, N_\mathrm{RF}$. Notice that  the SD beamformer in (\ref{frf2}) and (\ref{frf3}) includes the compensation of beam-squint via multiplying the phase terms by $\eta_m$. Next, we update the baseband beamformer in (\ref{fbb1}) such that our hybrid beamformer $\mathbf{F}_\mathrm{RF}^{(p)}\mathbf{F}_\mathrm{BB}^{(p)}[m]  $ resembles  the SD JSC beamformer  ${\mathbf{F}}_\mathrm{SC}^{(p)}[m]$ as much as possible. Hence, from (\ref{fbb1}), the updated baseband beamformer is computed as 
	\begin{align}
	\mathbf{F}_\mathrm{BB}^{(p)}[m] = \left(\mathbf{F}_\mathrm{RF}^{(p)}\right) ^\dagger \breve{\mathbf{F}}_\mathrm{RF}^{(p)}[m] \breve{\mathbf{F}}_\mathrm{BB}^{(p)}[m]. \label{fbbUpdated}
	\end{align}

	\subsubsection{Solve for $\widetilde{\mathbf{D}}^{(p)}$}
	Given   $\mathbf{F}_\mathrm{RF}^{(p)}$, $\widetilde{\mathbf{F}}_\mathrm{BB}^{(p)}$ and $\widetilde{\mathbf{F}}_\mathrm{SC}^{(p)} $,  the auxiliary variable $\widetilde{\mathbf{D}}^{(p)}$ can be optimized via
	\begin{align}
	\minimize_{\widetilde{\mathbf{D}}^{(p)} } \; \;& \|\mathbf{F}_\mathrm{RF}^{(p)}\widetilde{\mathbf{F}}_\mathrm{BB}^{(p)} - \widetilde{\mathbf{F}}_\mathrm{SC}^{(p)}   \|_\mathcal{F} \nonumber \\
	\subjectto & \widetilde{\mathbf{D}}^{(p)} \widetilde{\mathbf{D}}^{(p)^\textsf{H}} = \mathbf{I}_T,
	\end{align}
	\vspace{-12pt}
	which is called orthogonal Procrustes problem~\cite{elbir2023Mar_SPIM_ISAC}, and its solution is 
	\begin{align}
	\widetilde{\mathbf{D}}^{(p)} = {\mathbf{U}} \mathbf{I}_{T\times MN_\mathrm{ds}} {\mathbf{V}}, \label{D_update}
	\end{align}
	where ${\mathbf{U}} \boldsymbol{\Sigma} {\mathbf{V}} = {{\mathbf{F}}_\mathrm{S}^{(p)^\textsf{H}}}\mathbf{F}_\mathrm{RF}\widetilde{\mathbf{F}}_\mathrm{BB}^{(p)}$ is the SVD of the $T\times MN_\mathrm{ds}$ matrix $\frac{1}{1 - \varepsilon}{{\mathbf{F}}_\mathrm{S}^{(p)^\textsf{H}}} \left(\mathbf{F}_\mathrm{RF}^{(p)}\widetilde{\mathbf{F}}_\mathrm{BB}^{(p)} - \widetilde{\mathbf{F}}_\mathrm{SC}^{(p)} \right)$, and $\mathbf{I}_{T\times MN_\mathrm{ds}} = \left[\mathbf{I}_T | \mathbf{0}_{MN_\mathrm{ds}-T\times T}^\textsf{T} \right]$.

	
	In Algorithm~\ref{alg:HB}, the algorithmic steps of the proposed subarrayed beamforming approach are presented. Specifically, for a given subarray configuration index $p$, we first construct the subarray terms $\mathbf{F}_\mathrm{C}^{(p)}[m] $ and $\mathbf{F}_\mathrm{S}^{(p)} $ from the full array quantities. Then, for the first iteration, i.e., $j=1$,  the unknown variables are initialized as $\widetilde{\mathbf{D}}^{(p,j)} = \mathbf{I}_{T\times MN_\mathrm{ds}}$ and  $\mathbf{F}_\mathrm{RF}^{(p,j)}  = e^{\mathrm{j}\boldsymbol{\Psi}  }$, where $[\boldsymbol{\Psi}]_{k,n} \sim  \mathrm{uniform}([0,2\pi))$ for $k = 1,\cdots, K$ and $n = 1,\cdots, N_\mathrm{RF}$. During alternating optimization, firstly, the analog beamformer $\mathbf{F}_\mathrm{RF}^{(p)}$ is optimized in the steps 7-12. Then, the virtual SD analog beamformer $\breve{\mathbf{F}}_\mathrm{RF}^{(p)}[m]$ is computed, and the baseband beamformer is updated for BSC in steps 15-16. Finally, the auxiliary variable  $	{\mathbf{D}}^{(p)}[m]$ is updated in step 17. The hybrid beamformers are obtained when the algorithm converges~\cite{elbir2021JointRadarComm,hybridBFAltMin,manopt3_multiUser_Vandendorpe_Du2019Jul}.


	\begin{algorithm}[t]
		\begin{algorithmic}[1] 
			\caption{ \bf  ISAC Hybrid Beamforming With BSC }
			\Statex {\textbf{Input:} Subarray index $p$, $\overline{\mathbf{F}}_\mathrm{S}$, $\overline{\mathbf{H}}[m]$, $\eta_m$ and $\varepsilon$.  \label{alg:HB}}
			\State Compute $\overline{\mathbf{F}}_\mathrm{C}[m]$ from $\overline{\mathbf{H}}[m]$. 
			\State Construct $\mathbf{Q}_p$.
			\State  Initialize $\mathbf{F}_\mathrm{C}^{(p)}[m] =\mathbf{Q}_p \overline{\mathbf{F}}_\mathrm{C}[m] $, $\mathbf{F}_\mathrm{S}^{(p)} =\mathbf{Q}_p \overline{\mathbf{F}}_\mathrm{S}$,  $\mathbf{F}_\mathrm{RF}^{(p,j)} $ and  $\widetilde{\mathbf{D}}^{(p,j)} = \mathbf{I}_{T\times MN_\mathrm{ds}}$ for $j=1$.
			\State \textbf{while}   $\delta_j < \overline{\delta}$ \textbf{do}
			\State \indent ${\mathbf{F}}_\mathrm{SC}^{(p,j)}[m] \gets \varepsilon{\mathbf{F}}_\mathrm{C}^{(p)}[m] +  (1 - \varepsilon)\mathbf{F}_\mathrm{S}^{(p)}{\mathbf{D}}^{(p,j)} [m]  $.
			\State \indent  $\mathbf{f}_{j,i}^{(p)}\gets \mathrm{vec}\{\mathbf{F}_\mathrm{RF}^{(p,j)}\}$.
			\State \indent \textbf{while}
			\State \indent \indent Choose Armijo backtracking step size $\dot{a}_i$. 
			\State \indent \indent Compute the gradient $\nabla_\mathcal{R} \mathbf{f}_{j,i}^{(p)}$ in (\ref{gradR}).
			\State \indent \indent Update $\mathbf{f}_{j,i}^{(p)}$ from (\ref{f_i_MO_update}).
			\State \indent \indent $i \gets i + 1$.
			\State \indent \textbf{until} convergence
			\State \indent $\mathbf{F}_\mathrm{RF}^{(p,j)} \gets \mathrm{vec}^{-1}\{\mathbf{f}_{j,i}^{(p)}\}$.
			\State \indent Compute $\breve{\mathbf{F}}_\mathrm{RF}^{(p)}[m]$ from  (\ref{frf2}) and (\ref{frf3}).
			\State \indent Compute $\breve{\mathbf{F}}_\mathrm{BB}^{(p)}[m]$ from (\ref{fbb1}). 
			\State \indent Using $\breve{\mathbf{F}}_\mathrm{RF}^{(p)}[m]$ and $\breve{\mathbf{F}}_\mathrm{BB}^{(p)}[m]$, update $\mathbf{F}_\mathrm{BB}^{(p)}[m]$ as in \par \indent (\ref{fbbUpdated}) for beam-squint compensation.
			\State \indent Update $	{\mathbf{D}}^{(p)}[m]$ as in (\ref{D_update}).
			\State \indent $\delta_j = ||\mathbf{F}_\mathrm{RF}^{(p)}\widetilde{\mathbf{F}}_\mathrm{BB}^{(p)}- \widetilde{\mathbf{F}}_\mathrm{SC}^{(p)}    \|_\mathcal{F}  $
			\State \indent $j \gets j +1$. 
			\State \textbf{end} 
			\Statex \textbf{Return:} Hybrid beamformers $\mathbf{F}_\mathrm{RF}^{(p)}$ and $\mathbf{F}_\mathrm{BB}^{(p)}[m].$
		\end{algorithmic} 
	\end{algorithm}

	\subsection{Antenna Selection}
	Given the hybrid beamformers $\mathbf{F}_\mathrm{RF}^{(p)}$, $\mathbf{F}_\mathrm{BB}^{(p)}[m]$, we can write the SE when the $p$-th subarray is selected as
	\begin{align}
	\mathrm{SE}_p[m] &= \log_2 \big|\mathbf{I}_{N'} + \frac{1}{N_\mathrm{ds} \sigma^2} \mathbf{H}^{(p)}[m] \mathbf{F}_\mathrm{RF}^{(p)}\mathbf{F}_\mathrm{BB}^{(p)}[m] \nonumber \\
	&\hspace{40pt} \times \mathbf{F}_\mathrm{RF}^{(p)}{\mathbf{F}_\mathrm{BB}^{(p)}}^\textsf{H}[m] {\mathbf{H}^{(p)}}^\textsf{H}[m]       \big|, \label{SEp}
	\end{align}
	and the SE over all subcarriers is $\mathrm{SE}_p = \sum_{m = 1}^{M}\mathrm{SE}_p [m]$. Using (\ref{SEp}), the antenna selection problem is written as
	\begin{align}
	p^\star=&\arg \max_{p} \mathrm{SE}_p \nonumber \\
	\subjectto 	& \mathbf{H}^{(p)}[m] = \overline{\mathbf{H}}[m]\mathbf{Q}_p, \label{optAntSel}
	\end{align}
	where $p^\star$ represents the best subarray configuration maximizing the SE.

	{
		Now, we discuss the optimality of the best subarray configuration under the impact of beam-squint.  Define $\mathbf{u}_{p^\star}(\theta) = \mathbf{Q}_{p^\star}^\textsf{T}\mathbf{a}(\theta)$ as the $K\times 1$ beam-squint-free best subarray  steering vector corresponding to an arbitrary direction $\theta$. Then, it is clear that $\mathbf{u}_{p^\star}(\theta)$ achieves the highest SE in (\ref{optAntSel}) and  obtains the maximum array gain $ A_G(\Phi)$ for an arbitrary direction $\Phi$ if $\Phi = \theta$, i.e., $\theta	=\arg \maximize_{\Phi} A_G(\Phi),$	where
		\begin{align}
		A_G(\Phi) = { |\mathbf{u}_{p^\star}^\textsf{H}(\theta)\mathbf{u}_{p^\star}(\Phi)    |^2  }/{N^2}.
		\end{align}
		In the following theorem, we show that the array gain, and equivalently the SE~\cite{arrayGain_Capacity_Andersen2000Nov,arrayGain_diversity_Zheng2003May,elbir2023Mar_SPIM_ISAC}, is maximized only if two conditions are met, i.e., the best subarray configuration is selected and beam-squint is completely compensated.

		\begin{theorem}
			\label{lemma1}
			Define $\mathbf{u}_p(\vartheta_{m}) = \mathbf{Q}_{p}^\textsf{T}\mathbf{a}(\vartheta_{m})$ as the $K\times 1$ beam-squint-corrupted subarray steering vector   corresponding to an arbitrary direction $\theta$ and subcarrier $m\in \mathcal{M}$ as defined in (\ref{steringVec_aT}). Then,  $\mathbf{u}_p (\vartheta_{m})$ achieves the maximum array gain only if 
			\begin{align}
			\sin\vartheta_{m} = \eta_m \sin\theta,	\hspace{10pt}  p = p^\star  , 
			\end{align}
			where  $p^\star$ is the optimizer of (\ref{optAntSel}), and the array gain varying across the whole bandwidth is 
			\begin{align}
			A_G(\vartheta_m) = \frac{|\mathbf{u}_{p^\star}^\textsf{H}(\theta)  \mathbf{u}_{p}(\vartheta_{m})|^2}{N^2}. \label{arrayGain2}
			\end{align} 
		\end{theorem}

		\begin{IEEEproof}We first prove the condition $p=p^\star$ is required for the maximization of array gain in  (\ref{arrayGain2}).  Thus, we start by rewriting the array gain across the subcarriers in (\ref{arrayGain2})  as
			\begin{align}
			A_G(\vartheta_m)& = \frac{|\mathbf{u}_{p^\star}^\textsf{H}(\theta)  \mathbf{u}_{p}(\vartheta_{m})|^2}{N^2} \nonumber \\
			&= \frac{| \mathbf{a}^\textsf{H}(\theta)  \mathbf{Q}_{p^\star} \mathbf{Q}_{p}^\textsf{T} \mathbf{a}(\vartheta_{m})   |^2} {N^2} \nonumber \\
			&  = \frac{\mathrm{Trace}\{\mathbf{Q}_{p^\star} \mathbf{Q}_{p}^\textsf{T} \} | \mathbf{a}^\textsf{H}(\theta) \mathbf{a}(\vartheta_{m})  |^2}{N^2}, \label{arrayGain3}
			\end{align}
			where the first term in the nominator, i.e., $\mathrm{Trace}\{\mathbf{Q}_{p^\star} \mathbf{Q}_{p}^\textsf{T} \}$, is maximized as $	\max_{p} \mathrm{Trace}\{\mathbf{Q}_{p^\star} \mathbf{Q}_{p}^\textsf{T} \} =K$, which can be achieved only if $p= p^\star$ when $\mathbf{Q}_{p^\star} \mathbf{Q}_{p}^\textsf{T} = \mathbf{I}_K$.

			Now, we show that maximum array gain  is only achieved when $\sin \vartheta_{l,m} = \eta_m \sin \theta_l$. Thus, substituting  $\mathbf{Q}_{p^\star} \mathbf{Q}_{p}^\textsf{T} = \mathbf{I}_K$ in (\ref{arrayGain3}) yields
			\begin{align}
			&	A_G(\vartheta_m)  = \frac{ | \mathbf{a}^\textsf{H}(\theta) \mathbf{a}(\vartheta_{m})  |^2}{N^2} \nonumber \\
			&= \frac{1}{N^2} \left| \sum_{n_1 =1}^{N}  \sum_{n_2=1}^{N} e^{-\mathrm{j} \pi  \left( (n_1-1)\sin{\vartheta}_{m} - (n_2-1)\frac{\lambda_c\sin\theta}{\lambda_m}\right)    }   \right| ^2 \nonumber \\
			&= \left|\sum_{n = 0}^{N-1} \frac{e^{-\mathrm{j}2\pi n \bar{d} \left( \frac{\sin\vartheta_{m}}{\lambda_c} - \frac{\sin\theta}{\lambda_m}  \right)     }  }{N} \right|^2 
			\nonumber\\
			&
			=   \left|\sum_{n = 0}^{N-1} \frac{e^{-\mathrm{j}2\pi n \bar{d} \frac{(f_c \sin\vartheta_{m} - f_m\sin \theta) }{c}     }  }{N} \right|^2 \nonumber\\
			&= \left| \frac{1 - e^{-\mathrm{j}2\pi N\bar{d} \frac{(f_c \sin\vartheta_{m}- f_m\sin\theta)}{c}    }}{N (1 - e^{-\mathrm{j}2\pi \bar{d}\frac{(f_c \sin\vartheta_{m} - f_m\sin\theta)}{c} ) }}   \right|^2 \hspace{-3pt}
			\nonumber\\
			&
			= \left| \frac{\sin (\pi N \mu_m )}{N\sin (\pi \mu_m )}    \right|^2 \hspace{-3pt}  = |\zeta( \mu_m )|^2, \label{arrayGain}  
			\end{align}
			where    $\mu_m = \bar{d}\frac{(f_c \sin\vartheta_{m}- f_m\sin\theta)}{c}   $. Due to the power focusing capability of the Drichlet sinc function $\zeta (\mu_m)$ in  (\ref{arrayGain}), array gain  is focused only on a small portion of the beamspace, and it substantially reduces across the subcarriers as $|f_m - f_c|$ increases. Furthermore, $|\zeta( \mu_m )|^2$ gives peak when $\mu_m = 0$, i.e.,  $f_c\sin \vartheta_{m} - f_m\sin\theta= 0$, which yields $ \sin\vartheta_{m} = \eta_m \sin\theta$.
		\end{IEEEproof}
		
	}

	The problem in (\ref{optAntSel}) requires to visit all possible subarray configurations, which can be computationally prohibitive and consume too much memory resources, especially when $N$ is large, e.g., $N \geq 32$. To reduce this cost, we propose low complexity algorithms in the following.

	\begin{figure}[t]
		\centering
		{\includegraphics[draft=false,width=.96\columnwidth]{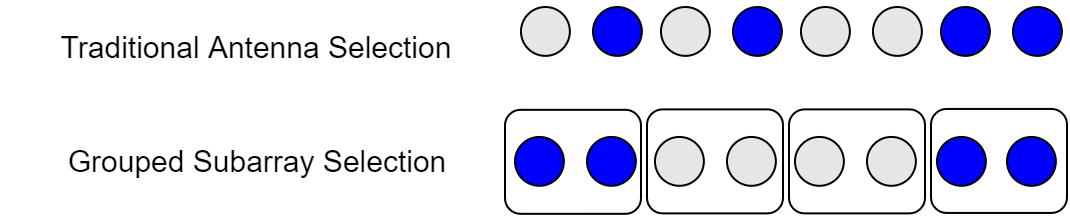}} 
		\caption{An example of selecting $4$  out of $8$ antennas for traditional antenna selection and GSS, wherein the antennas are picked individually and in groups of size $G$, respectively. 
		}
		\label{fig_AntennaSelectionSchemes}
	\end{figure}
	%
	%
	%

	\subsubsection{Grouped Subarray Selection}
	In order to reduce the computational cost involved in the solution of (\ref{optAntSel}), we propose GSS strategy to lower the  number of possible subarray configurations.
	In GSS, the whole array is divided into $N_G = \frac{N}{G}$  disjoint groups, each of which includes $G$ consecutive antennas as illustrated in Fig.~\ref{fig_AntennaSelectionSchemes}.  Thus, the antenna selection problem with GSS reduces to selecting $K_G = \frac{K}{G}$ groups out of $N_G$, and the number of possible subarray configurations with GSS is $P_G =  \frac{N_G!}{K_G! (N_G - K_G)!  }$.  Then, the set of all possible subarray configurations with GSS is $\mathcal{Q}_G = \{\mathbf{Q}_1,\cdots, \mathbf{Q}_{P_G}   \}$, where $\mathbf{Q}_{p_G}$ represents the antenna selection matrix for the $p_G$-th configuration, $p_G = 1,\cdots, P_G$.

	The grouped array  structure significantly reduces the number of subarray configurations, even for small number of grouped antennas, e.g., $G=2$. As an example, consider the scenario when $N = 64$ and $K = 32$. Then, the number of subarray configurations is $P = 1.83\times 10^{18}$, which reduces to approximately $P_2 = 6\times 10^{9}$ ($3\times 10^{9}$ times fewer) and  $P_4 = 12870$ ($1.4\times 10^{14}$ times fewer) for $G = 2$ and $G= 4$, respectively.

	\begin{algorithm}[t]
		\begin{algorithmic}[1] 
			\caption{ \bf  ISAC Beamforming \& Antenna Selection }
			\Statex {\textbf{Input:} $N$, $K$, $\overline{\mathbf{F}}_\mathrm{S}$, $\overline{\mathbf{H}}[m]$, $\forall m$, $P$, $B$. \label{alg:Search}}
			\State Initialize $\mathrm{SE}_0 = 0$, 
			\State  \textbf{for} $b= 1,\cdots, B$ \textbf{do}
			\State \indent Construct $\mathcal{Q}^b$.
			\State \indent \textbf{for} $p \in \{\frac{P(b-1)}{B}+1,\cdots,  \frac{Pb}{B}\}$ \textbf{do}
			\State \indent \indent Construct $\mathbf{Q}_p$.
			\State \indent \indent  $\mathbf{H}^{(p)}[m] \hspace{-2pt} = \hspace{-2pt} \overline{\mathbf{H}}[m]\mathbf{Q}_p$, $\forall m$.
			\State \indent \indent Compute $\mathbf{F}_\mathrm{RF}^{(p)}$ and $\mathbf{F}_\mathrm{BB}^{(p)}[m]$ from Algorithm~\ref{alg:HB}.
			\State \indent \indent Compute $\mathrm{SE}_p$ from (\ref{SEp}).
			\State \indent \textbf{end}
			\State \indent  Solve for $q_b^\star$ as $q_b^\star = \arg \max_{q_b} \mathrm{SE}_{q_b}$.
			\State \indent  $q_b^\star \gets  \left\{\begin{array}{cc}
			q_b^\star,& \textrm{if } \mathrm{SE}_{q_{b-1}^\star } < \mathrm{SE}_{q_b^\star}, \\
			q_{b-1}^\star, & \textrm{otherwise}
			\end{array}\right..$
			\State \indent  Save $q_b^\star$ and clear the memory of $\mathbf{F}_\mathrm{RF}^{(p)}$, $\mathbf{F}_\mathrm{BB}^{(p)}[m]$ and \par \indent     $\mathbf{H}^{(p)}[m]$, $\forall m,p$.
			\State \textbf{end} \textbf{for}
			\State \textbf{Return} Best subarray configuration index $q^\star = q_B^\star$, and the hybrid beamformers $\mathbf{F}_\mathrm{RF}^{(q^\star)}$ and $\mathbf{F}_\mathrm{BB}^{(q^\star)}[m]$.
		\end{algorithmic} 
	\end{algorithm}
	
	\subsubsection{Sequential Search Algorithm}
	During the computation of $\mathrm{SE}_p$ for very large number of $P$, the computation platform requires very large amount of memory to save the variables whose dimensions are proportional to $P$. To efficiently use the memory, we devise a sequential search algorithm, wherein $\mathcal{Q}$ is partitioned into $B$ disjoint blocks as $\mathcal{Q} = \cup_{b = 1}^B \mathcal{Q}^b $, where $\mathcal{Q}^b =  \{\mathcal{Q}_{\frac{P(b-1)}{B}+1},\cdots, \mathcal{Q}_{\frac{Pb}{B}} \}$. Then, (\ref{optAntSel}) is sequentially solved such that the variables, e.g.,  $\mathbf{F}_\mathrm{RF}^{(p)}\mathbf{F}_\mathrm{BB}^{(p)}[m]$ and $\mathbf{H}^{(p)}[m]$ for $p \in \{\frac{P(b-1)}{B}+1,\cdots,  \frac{Pb}{B}\}$, are removed from the memory after computation at the block $b$, instead of storing all the data in the memory. As a result, the computational platform requires approximately $B$ times less memory.

	{In Algorithm~\ref{alg:Search}, we present the algorithmic steps of the proposed approach for ISAC hybrid beamformer design with antenna selection. First, we compute $\mathcal{Q}^b$, i.e., the set of subarray configurations for the $b$-th block. Then, we design the hybrid beamformers and the cost (i.e., SE) is computed. Next, the computed costs of two consecutive blocks (i.e., $\mathrm{SE}_{q_{b-1}^\star }$ and $ \mathrm{SE}_{q_b^\star}$) are compared, and the unnecessary variables are removed from the memory. Following this strategy, $\forall b$, the best subarray index $q^\star$ and the hybrid beamformers $\mathbf{F}_\mathrm{RF}^{(q^\star)}$ and $\mathbf{F}_\mathrm{BB}^{(q^\star)}[m]$.

	}

	%
	%

	\subsection{Learning-Based  Antenna Selection}
	\label{sec:ML_AS}
\begin{figure}[t]
		\centering
		{\includegraphics[draft=false,width=\columnwidth]{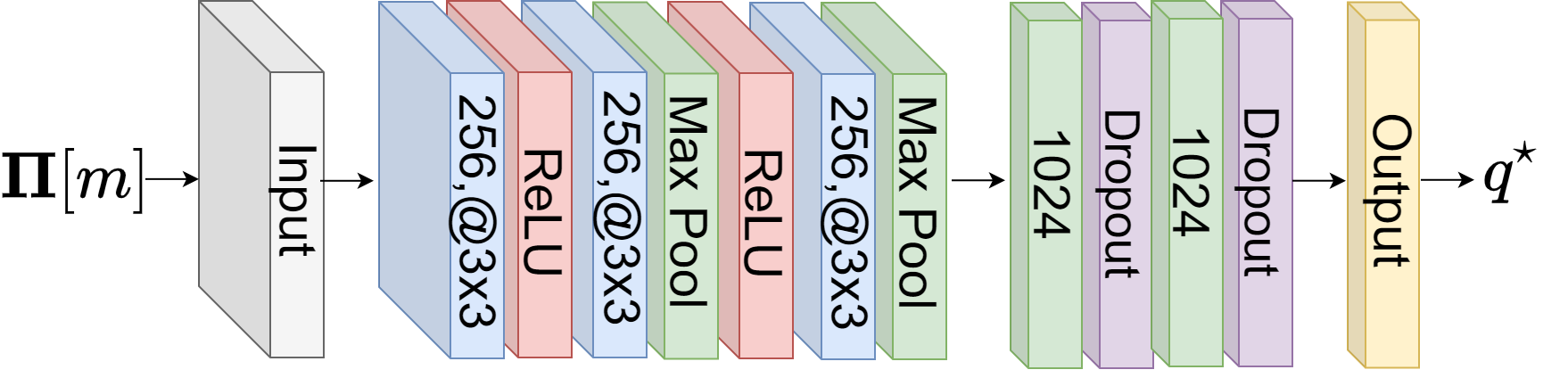}} 
		\caption{The CNN architecture for antenna selection. 
		}
		\label{fig_CNN}
	\end{figure}

	Due to the combinatorial nature of antenna selection problem, it is preferable to formulate the problem as a classification problem, wherein each subarray configuration is regarded as a class. Thus, we design a classification model with a CNN architecture as shown in Fig.~\ref{fig_CNN}. Define $\mathcal{D} = \{\mathcal{D}_1, \cdots, \mathcal{D}_\textsf{I} \}$ as the training dataset, where  $\mathcal{D}_i = (\mathcal{I}_{i},\mathcal{O}_i)$ denotes the $i$-th input and output data for $i = 1,\cdots, \textsf{I}$. The input of the CNN is formed from the combination of communications (channel matrix) and sensing (received target responses) data $\boldsymbol{\Pi}[m] \in \mathbb{C}^{N \times (N' + T )} $ as
	\begin{align}
	\boldsymbol{\Pi}[m] = \left[\underbrace{\overline{\mathbf{H}}^\textsf{T}[m]}_{\mathrm{Communications}}, \underbrace{\overline{\mathbf{F}}_\mathrm{S}}_{\mathrm{Sensing}}  \right]^\textsf{T}.
	\end{align}
	Define $\boldsymbol{\Pi}^{(i)} [m]\in \mathbb{C}^{N \times (N' + T )}$ as the generated data for the $i$-th sample of the dataset. Then, the input includes the real and imaginary parts of $\boldsymbol{\Pi}^{(i)}[m]$ as $\mathcal{X}_{i,1} = \operatorname{Re}\{ \boldsymbol{\Pi}^{(i)} [m]\}$ and $\mathcal{X}_{i,2} = \operatorname{Im}\{ \boldsymbol{\Pi}^{(i)}[m] \}$, respectively. Thus, $\mathcal{X}_{i}$ is a ``two-channel" real-valued tensor variable with the size of $N \times (N' + T )\times 2$. Furthermore, the output of the $i$-th sample is represented the best subarray index obtained from  Algorithm~\ref{alg:Search} as $\mathcal{O}_i = q_{(i)}^{\star}$. As a result, the output includes possible subarray configurations as $\mathcal{O}_i \in \mathcal{Q}$. Let $\boldsymbol{\theta}\in \mathbb{R}^{U}$ denote the learnable parameters of the CNN. Then, the learning model aims to construct  the non-linear mapping between the input $\mathcal{I}_i$ and the output label $\mathcal{O}_i$  as  $\mathcal{F}(\boldsymbol{\theta}, \mathcal{I}_i) \rightarrow \mathcal{O}_i$.

	The CNN architecture for antenna selection  has 13 layers as shown in Fig.~\ref{fig_CNN}. The first layer is the input layer with the size of $N \times (N' + T )\times 2$. The $\{2,4,7\}$th layers are convolutional layers  with $256$ filter and kernel size of $3\times 3$. The third and the sixth layers are rectified linear unit layers performing nonlinear feature mapping of $f_\mathrm{ReLU}(x) = \max(0,x)$ for its input $x$. The fifth and the eighth layers are composed of pooling layers, which reduce the dimension by 2. There are fully connected layers, each of which is followed by a dropout layer with probability of $0.5$, at the ninth and eleventh layers with 1024 units. The output layer is a classification layer with softmax function and $P$ classes, each of which corresponds to a distinct subarray configuration.

	\section{Numerical Experiments}
	\label{sec:sim}
	We evaluate the performance of the proposed approach via several experiments. During simulations, the target and user path directions are drawn uniform randomly as $\Phi_k,\phi_l,\theta_l \in [30^\circ, 150^\circ]$. We conducted $500$ Monte Carlo trials, and presented the averaged results. We assumed that there are $T=3$ point targets and a single user with $L=3$ NLoS paths. We select $N' = 16$,  $ N_\mathrm{RF} = 8$, $M=16$ and the number of data snapshots $T_\mathrm{S}=256$.

	\begin{figure}[t]
		\centering
		\subfloat[$N=32$, $K=8$]{\includegraphics[draft=false,width=\columnwidth]{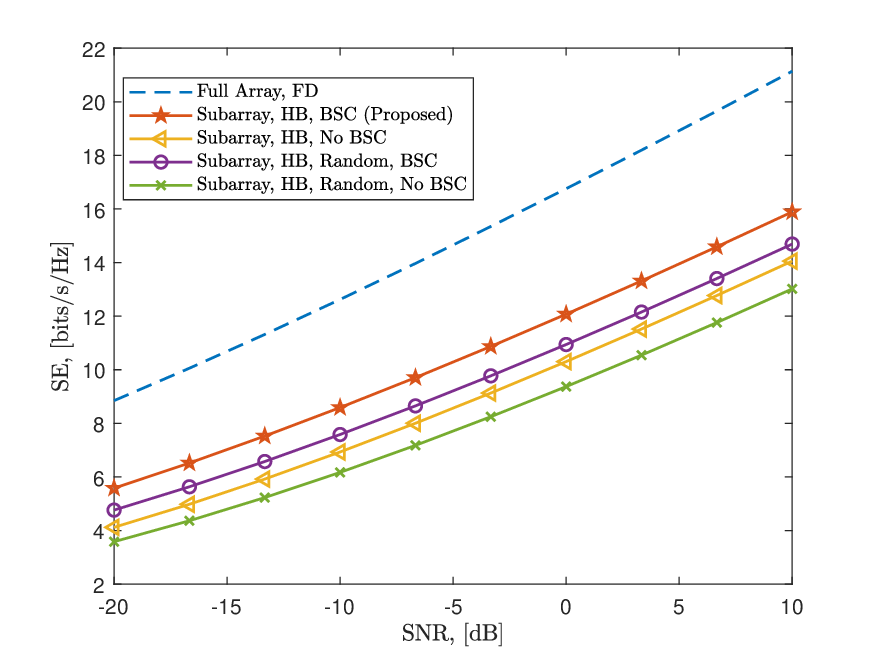}}  \\
		\subfloat[$N=128$, $K=8$]{\includegraphics[draft=false,width=\columnwidth]{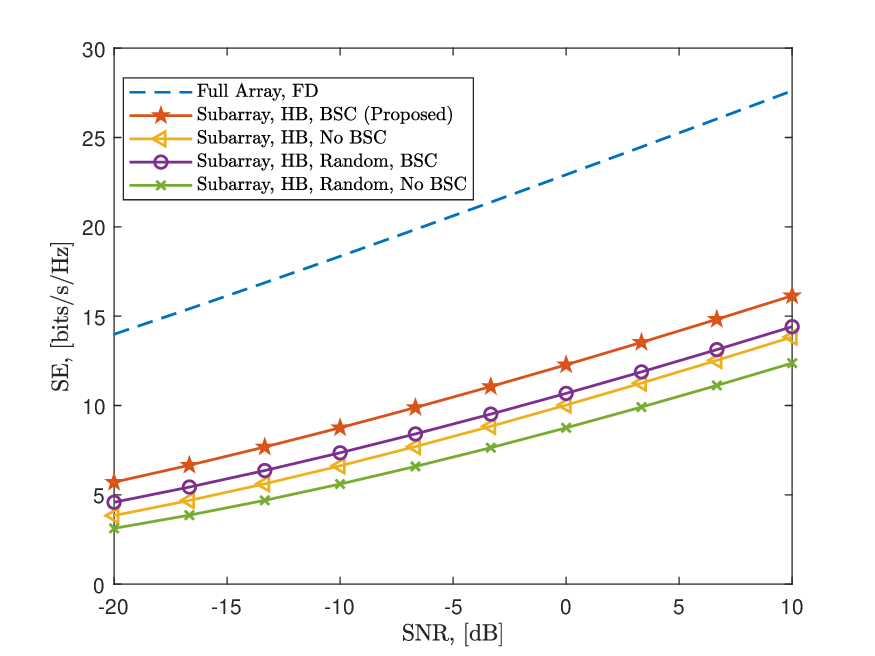}} 
		\caption{SE performance versus SNR for (a) $N=32$ and (b) $N=128$ when $K=8$, $G=4$ and $\varepsilon=0.5$.
		}
		\label{fig_SE_SNR}
	\end{figure}

	\begin{figure}[t]
		\centering
		{\includegraphics[draft=false,width=\columnwidth]{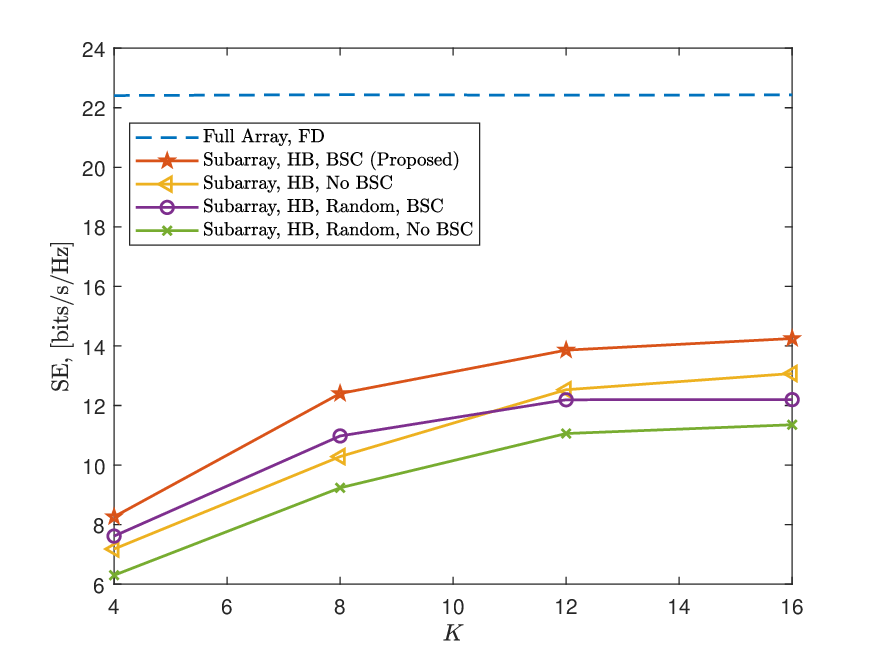}} 
		\caption{SE versus number of selected antennas $K$  when $N=64$, $G=4$, $\varepsilon=0.5$ and $\mathrm{SNR}=0$ dB.
		}
		\label{fig_SE_K}
	\end{figure}
	
	\begin{figure}[t]
		\centering
		{\includegraphics[draft=false,width=\columnwidth]{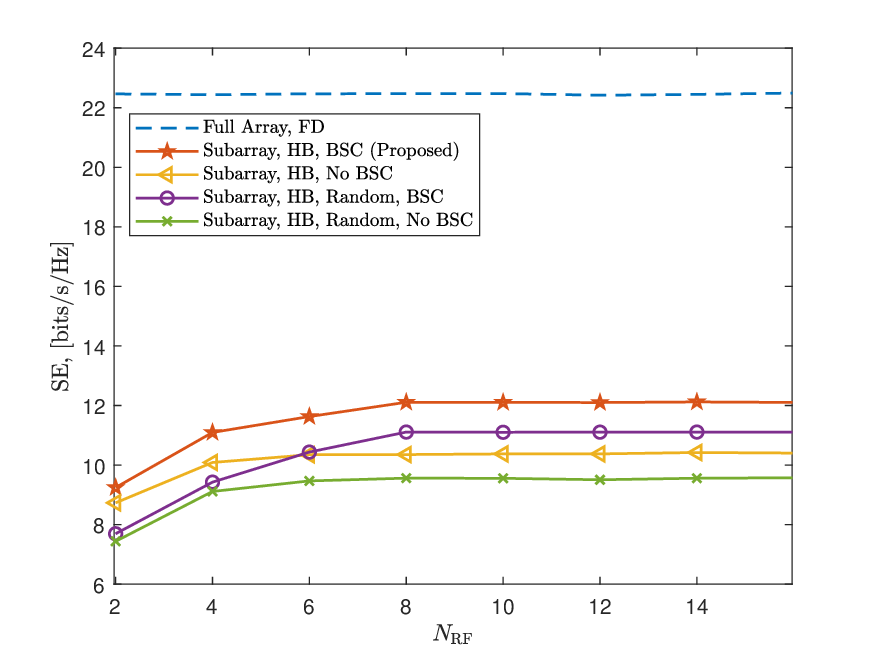}} 
		\caption{SE and number of RF chains when $N=128$, $K=8$, $G=4$, $\varepsilon=0.5$ and $\mathrm{SNR}=0$ dB.
		}
		\label{fig_SE_NRF}
	\end{figure}

	\begin{figure}[t]
		\centering
		\subfloat[]{\includegraphics[draft=false,width=.42\textwidth]{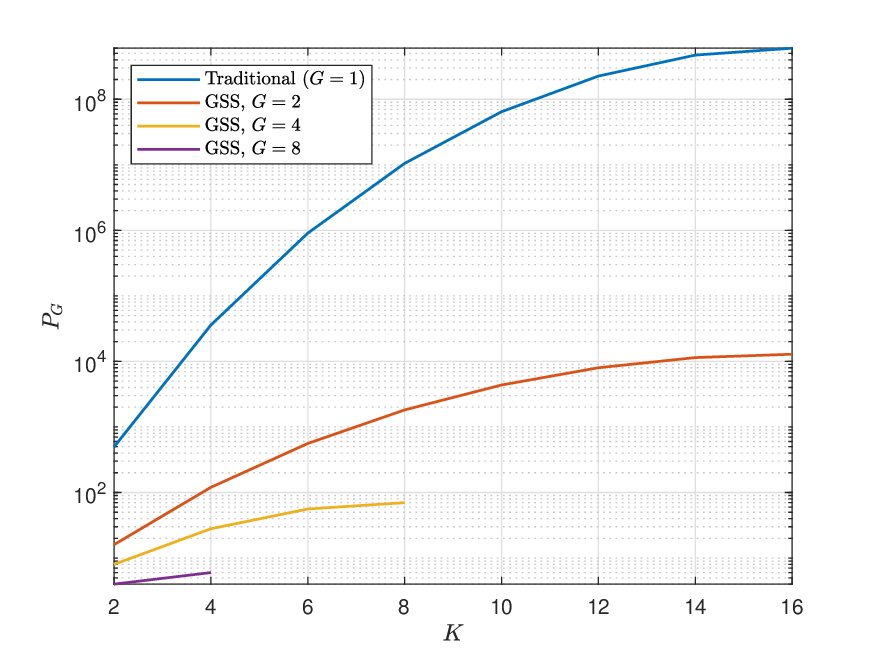}} \\
		\subfloat[]{\includegraphics[draft=false,width=.42\textwidth]{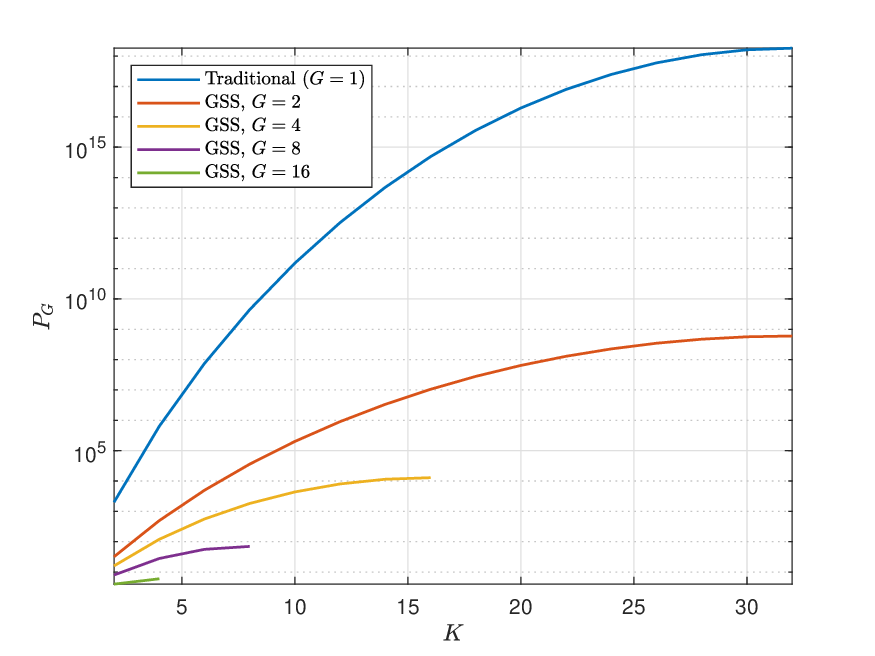}} \\
		\subfloat[]{\includegraphics[draft=false,width=.42\textwidth]{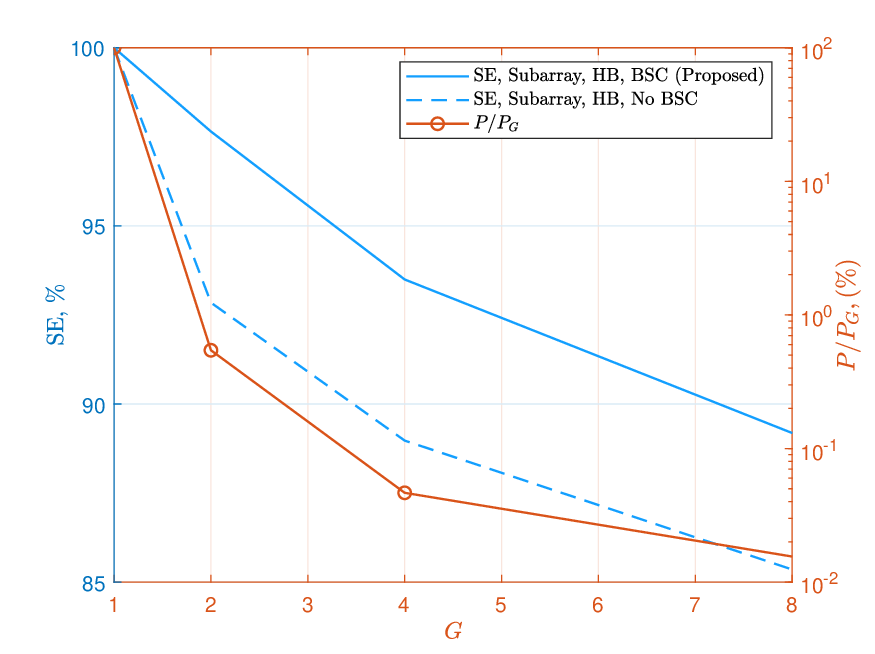}} 
		\caption{Number of possible subarray configurations for traditional antenna selection and GSS when (a) $N=32$ and (b) $N=64$. (c) SE and number of subarray configurations versus $G$ when $N=16$, $K=8$, $\varepsilon=0.5$ and $\mathrm{SNR}=0$ dB.
		}
		\label{fig_SE_G}
	\end{figure}
	
	Fig.~\ref{fig_SE_SNR} shows the SE performance of the proposed antenna selection and hybrid beamforming approach based on GSS and BSC for $N=32$ and $N=128$ when $K=8$ and $G=4$. The performance of the  fully digital (FD) full array beamformer (i.e., $\overline{\mathbf{F}}_\mathrm{SC} [m] = \varepsilon\overline{\mathbf{F}}_\mathrm{C}[m] + (1-\varepsilon) \overline{\mathbf{F}}_\mathrm{S}\mathbf{D}[m]$) is regarded as benchmark for the subarrayed hybrid beamformer, i.e., $\mathbf{F}_\mathrm{RF}^{(q^\star)}\mathbf{F}_\mathrm{BB}^{(q^\star)}[m]$. The beamforming and antenna selection are performed as described in Algorithm~\ref{alg:HB} and Algorithm~\ref{alg:Search}, respectively. We see from Fig.~\ref{fig_SE_SNR} that our BSC approach provides a significant SE improvement for hybrid beamforming. In order to evaluate  the performance of antenna selection accuracy, we also present the hybrid beamforming performance of the randomly selected subarrays with and without BSC. While the former design involves a BSC beamformer for a random subarray configuration, the latter does not involve BSC for the same randomly selected subarray.   We observed that the randomly selected beamformer with BSC performs better than that of optimized subarray without BSC for different array sizes. In other words, compensating for beam-squint plays more crucial role than optimizing for best subarray.

	Next, we present the SE performance in Fig.~\ref{fig_SE_K} with respect to the number of selected antennas $K$ when the transmit array size is fixed as $N = 64$ with $G=4$.  {We aim to generate $T + L=6$ disjoint beams towards $T=3$ targets and $L=3$ user path directions. Thus, we see that the SE is improved as $K$ increases, especially for  $K\geq T + L = 6$,  thanks to achieving higher beamforming gain.} We also observe that optimized subarray without BSC achieves higher SE than the random subarray with BSC as $K\geq10$. This suggests that implementing BSC in the case of the random subarray can improve the SE up to a certain extent. Furthermore, increasing the subarray size can potentially enhance the array gain, thereby mitigating the SE loss caused by beam-squint. 

	\begin{figure}[t!]
		\centering
		{\includegraphics[draft=false,width=\columnwidth]{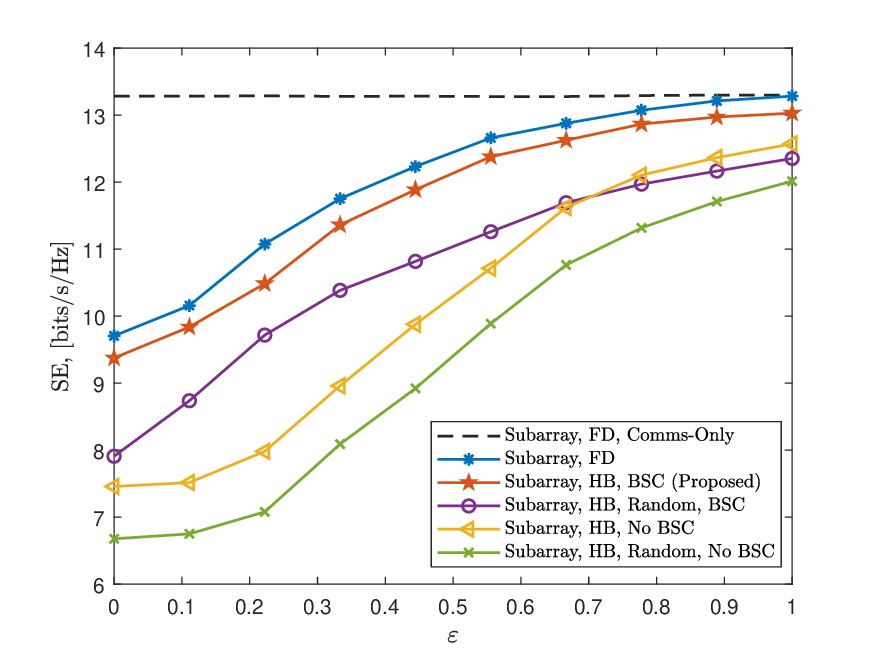}} 
		\caption{SE versus trade-off parameter $\varepsilon$ for $N=32$, $K=8$, $G=4$ and $\mathrm{SNR}=0$ dB.
		}
		\label{fig_SE_eta}
	\end{figure}

	The SE performance is presented in  Fig.~\ref{fig_SE_NRF} versus the number of RF chains $N_\mathrm{RF}$ for $N=128$ and $K=8$. As $N_\mathrm{RF} \geq 6$, higher SE is achieved by the randomly selected subarray even without BSC. In contrast, the SE performance of the BSC algorithms is reduced for $N_\mathrm{RF} < 8$. This is because the term $\left(\mathbf{F}_\mathrm{RF}^{(q^\star)}\right) ^\dagger \breve{\mathbf{F}}_\mathrm{RF}^{(q^\star)}[m]$ in (\ref{fbbUpdated}) becomes full column rank as $N_\mathrm{RF}\rightarrow N$, hence provides a better mapping from $	 \breve{\mathbf{F}}_\mathrm{BB}^{(q^\star)}[m]$ to $\mathbf{F}_\mathrm{BB}^{(q^\star)}[m]$. Nevertheless, our  optimized subarray with BSC exhibits  approximately $9\%$ and $27\%$  higher SE as compared to random subarray with and without BSC, respectively.

	In Fig.~\ref{fig_SE_G}, we evaluate the performance of our GSS approach in terms of number of subarray candidates as well as the loss in SE. Fig.~\ref{fig_SE_G}(a-b) shows that the number of possible subarray configurations significantly reduces with the slight increase of $G$ while the SE of the selected subarray degrades. This is because the antenna selection algorithm with GSS  visits only the subarray configurations in $\mathcal{Q}_G$ while leaving out the remaining candidates, which are in the set $\mathcal{Q} \backslash \mathcal{Q}_G $. Nevertheless, our GSS approach with BSC-based hybrid beamforming achieves significantly low complexities while maintaining a satisfactory SE performance as shown in Fig.~\ref{fig_SE_G}(c). In particular, when $G=4$, the number of subarray candidates are reduced about $95\%$ while our BSC approach yields only $6\%$ loss in SE, whereas this loss is approximately $11\%$ for the subarray beamforming without BSC.

	The trade-off between communications and sensing is evaluated in Fig.~\ref{fig_SE_eta} for $\varepsilon \in [0,1]$, $N=32$, $K=8$, and $G=4$. As a benchmark, the ISAC hybrid beamforming approach is also included with the FD communications-only beamformer (i.e., $\mathbf{F}_\mathrm{C}^{(q^\star)}[m]$) as well as the FD beamformer given in (\ref{FSC}), which depends on $\varepsilon$. The FD ISAC beamformer attains the FD communications-only beamformer for $\varepsilon = 1$ as expected while its performance degrades as $\varepsilon \rightarrow 0$ as the trade-off becomes sensing-weighted. Furthermore, we observe that our beamformer with BSC closely follows the FD beamformer with significant improvement compared to beam-squint corrupted and randomly selected designs.

	\begin{figure}[t]
		\centering
		\subfloat[]{\includegraphics[draft=false,width=\columnwidth]{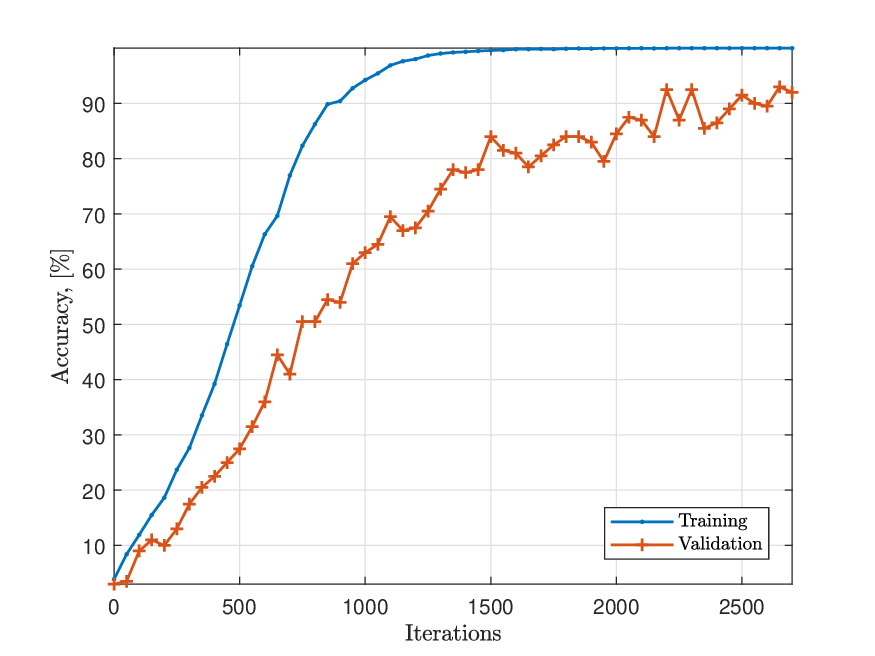}} \\
		\subfloat[]{\includegraphics[draft=false,width=\columnwidth]{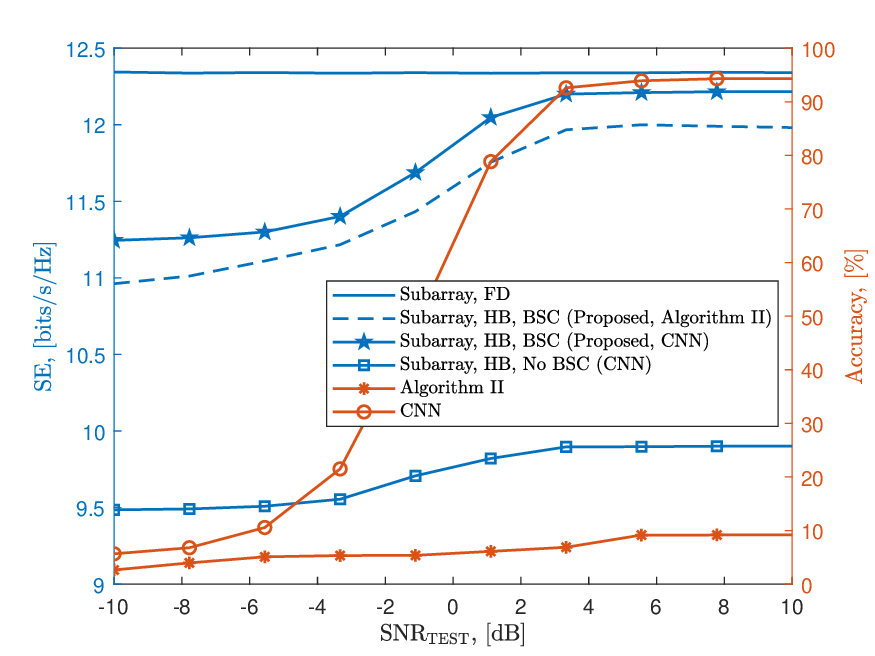}}
		\caption{Performance of the learning-based antenna selection. (a) training and validation accuracy during training. (b) SE performance versus $\mathrm{SNR}_\mathrm{TEST}$ when $N=32$,  $K=8$, $G=4$ and $\varepsilon=0.5$.
		}
		\label{fig_learning}
	\end{figure}

	Now, we evaluate the performance of our CNN model for antenna selection. The training dataset $\mathcal{D}$ is generated for $\textsf{I}_1 = 1000$ different data realizations, i.e., $\boldsymbol{\Pi}[m]$. Then, for each realized data sample,  synthetic noise is added onto the input data in order to make the model robust against corruptions and imperfections~\cite{elbir_aFamily_onlineDL_CE_HB_Elbir2021Dec,elbirQuantizedCNN2019}. Specifically, $\textsf{I}_2 = 100$ noise realizations are obtained for three different SNR values as $\mathrm{SNR}_\mathrm{TRAIN} \in \{15, 20,25\}$ dB. This process is repeated for each data realizations $i_1 = 1,\cdots, \textsf{I}_\mathrm{1}$ and $m =1,\cdots, M$ when $T=3$, $N'=16$, $N=32$, $K=8$ and $G=4$. As a result, the whole dataset includes $\textsf{I} = \textsf{I}_1 \cdot \textsf{I}_2 \cdot M \cdot 3= 480,000$ samples, each of which is of size $32\times 19\times 2$. The whole data set is divided into $30\%$ and $70\%$ two parts for validation and training, respectively. The validation dataset is then used for testing the CNN model after it is corrupted by synthetic noise with the SNR of $\mathrm{SNR}_\mathrm{TEST}$. Stochastic gradient descent algorithm with momentum of 0.9 is adopted during training, for which  the cross-entropy loss function is used for classification as
	\begin{align}
	\mathrm{ce} =  -\frac{1}{\textsf{I}} \sum_{i = 1}^{\textsf{I}} \sum_{p = 1}^{P} \bigg( \omega_{i,p}  \ln (\kappa_{i,p}) \nonumber  + (1- \omega_{i,p}) \ln (1-\kappa_{i,p}) \bigg)
	\end{align}
	where $\omega_{i,p}$ and $\kappa_{i,p}$ are the input-output pair of the classification layer defined for the $i$-th data sample and $p$-th class. The learning rate for training is set as 0.01 and it is reduced by half after each 500 iterations. The performance of the CNN model is presented in Fig.\ref{fig_learning}. Specifically, Fig.~\ref{fig_learning}(a) shows the classification (subarray selection) accuracies for validation and training. We can see that the CNN model successfully learns the training data while the accuracy for validation data is slightly less with approximately 90\% accuracy. This is because the validation data is not used during training. {We also present the antenna selection accuracy and the SE performance of the selected subarrays after employing Algorithm~\ref{alg:Search} and CNN in Fig.~\ref{fig_learning}(b). In this setup, the validation data is corrupted by synthetic noise defined by $\mathrm{SNR}_\mathrm{TEST}$ in order to assess the robustness of the proposed CNN model and Algorithm~\ref{alg:Search}, for which the corrupted data (i.e., $\boldsymbol{\Pi} [m]$) is used as input, and the corresponding hybrid beamformers are estimated. Then, the true channel data is used to compute the SE with the resulting hybrid beamformers. Our CNN model achieves up to 95\% antenna selection accuracy for the corrupted input data with $10$ dB noise while the model-based approach in Algorithm~\ref{alg:Search} is unable to provide accurate antenna selection results (approximately $10\%$) due to the corruptions in the data. This observation clearly shows the advantage of using learning-models in imperfect scenarios. Specifically, the CNN accounts for the imperfections in the data and yields the accurate subarray selection performance thanks to training with noisy communications and sensing data for robustness. When we compare the SE of model-based and learning-based approaches (in the left axis), we can see that the corrupted input data causes selecting inaccurate subarray indices, thereby leading to poor SE performance. Nevertheless,  CNN-based antenna selection yields  (about $2\%$) higher SE than the model-based approach in Algorithm~\ref{alg:Search} in the presence of imperfect communications and sensing data. 
		
	}

	\begin{table}[t!]
		\caption{Computation Times (In Seconds) }
		\footnotesize
		\label{tableComputationTimes1}
		\centering
		\begin{tabular}{ccc}
			\hline 
			\hline 
			& Model-based & Learning-based \\
			\hline
			$K=2$ & $0.287$ & $0.037$  \\
			\hline
			$K=3$ & $1.876$ & $0.037$  \\
			\hline
			$K=4$ & $22.270$ & $0.037$  \\
			\hline
			$K=5$ & $67.946$ & $0.037$  \\
			\hline
			$K=6$ & $116.463$ & $0.037$  \\
			\hline
			$K=7$ & $184.378$ & $0.037$  \\
			\hline
			$K=8$ & $269.441$ & $0.038$  \\
			\hline
			\hline 
		\end{tabular}
	\end{table}

	\begin{table}[t]
		\caption{Computation Times (In Seconds) For   Selecting  $K=8$ out of $N=16$ Antennas }
		\footnotesize
		\label{tableComputationTimes2}
		\centering
		\begin{tabular}{cccc}
			\hline 
			\hline 
			&$G=1$ &  $G=2$ & $G=4$ \\
			\hline
			Model-based &$269.441$ & $20.865$ & $0.328$  \\
			\hline
			Learning-based &$0.037$ & $0.037$ & $0.037$  \\
			\hline
			\hline 
		\end{tabular}
	\end{table}

	{Finally, we present the computation times (in seconds) for joint antenna selection and hybrid beamforming approach in Table~\ref{tableComputationTimes1} and Table~\ref{tableComputationTimes2}. In particular, Table~\ref{tableComputationTimes1} shows the computation times of model-based (Algorithm~\ref{alg:Search}) and learning-based approaches for $K\in \{2,\cdots, 8\}$, $N=16$ and $G=1$. While the complexity of the model-based approach grows geometrically, the learning-based approach with the CNN model  in Fig.~\ref{fig_CNN} enjoys significantly low computation times. The fast implementation of CNN is attributed to employing parallel processing tools e.g., graphical computation units (GPUs).  We also present the computation times with respect to the group size $G$ in Table~\ref{tableComputationTimes2} when $N=16$ and $K=8$. Note that the time complexity when $G=1$ corresponds to the traditional antenna selection whereas $G>1$ is for the proposed GSS strategy. The results are in accordance with Fig.~\ref{fig_SE_G} where a performance analysis with respect to $P$ and  $G$ are presented. We see that our  GSS approach approximately $11$ and $630$ times faster than traditional antenna selection for $G=2$ and $G=4$, respectively. Furthermore, the time complexity of learning-based antenna selection with the CNN model exhibits approximately the same amount of time for $G\in \{1,2,4\}$ while providing a significant reduction ($\sim 7000$ times faster) for computing the best subarray index as compared to the model-based approach.     }

	\section{Summary}
	\label{sec:summary}
	In this paper, we investigated the antenna selection problem in the presence of beam-squint for THz-ISAC hybrid beamforming. We have shown that the impact of beam-squint on antenna selection causes significant performance loss in terms of SE due to selecting inaccurate subarrays. The compensation for beam-squint during hybrid beamforming design is achieved via manifold optimization integrated with the proposed BSC algorithm. In particular, BSC provides a solution via updating the baseband beamformers by taking into account the distortions in the analog domain due to beam-squint. Specifically, BSC exhibits approximately $15\%$ improvement in terms of SE without requiring additional hardware components. 	In order to solve the joint antenna selection and hybrid beamforming, low complexity algorithms are proposed to reduce the number of possible subarray configurations. These include sequential search algorithm to reduce the memory usage during the computation of the subarray variables and GSS to reduce the number of possible subarray configurations via selecting the antennas in small groups. It is shown that the proposed GSS approach provides approximately $95\%$ reduction in terms of computational complexity while maintaining satisfactory performance with about $6\%$ SE loss.

	
	\footnotesize
	\bibliographystyle{IEEEtran}
	\bibliography{IEEEabrv,references_123}

\end{document}